\documentclass[lettersize,journal]{IEEEtran}
\usepackage{amsmath,amssymb,amsfonts}
\usepackage[utf8]{inputenc}
\usepackage{algorithmic}
\usepackage{algorithm}
\usepackage{array}
\usepackage[caption=false,font=scriptsize,labelfont=sf,textfont=sf]{subfig}
\usepackage{textcomp}
\usepackage{stfloats}
\usepackage{url}
\usepackage{verbatim}
\usepackage{graphicx}
\usepackage{cite}
\usepackage{amssymb}
\usepackage{setspace}
\usepackage{lineno} % 这个包就是latex自带的添加行号的包了
\usepackage{cases}
\usepackage{subfloat}
\usepackage{float}  %设置图片浮动位置的宏包
\usepackage{titlesec}
\usepackage{textcomp}
\usepackage{bm,bbold}
\usepackage{xcolor}
\usepackage{siunitx}
\newcommand\bw{\ensuremath{{\bm w}}}

\newcommand\bW{\ensuremath{{\bold W}}}

\newcommand\bq{\ensuremath{{\bm q}}}
\newcommand\bx{\ensuremath{{\bm x}}}

\newcommand\bG{\ensuremath{{\bold G}}}
\newcommand\bh{\ensuremath{{\bm h}}}
\newcommand\bH{\ensuremath{{\bold H}}}

\newcommand\be{\ensuremath{{\bm e}}}

\newcommand\bp{\ensuremath{{\bm p}}}

\newcommand\bC{\ensuremath{{\bold C}}}
\newcommand\bc{\ensuremath{{\bm c}}}
\newcommand\ba{\ensuremath{{\bm a}}}

\newcommand\bA{\ensuremath{{\bold A}}}

\newcommand\bb{\ensuremath{{\bm b}}}
\newcommand\bg{\ensuremath{{\bm g}}}
\newcommand\bB{\ensuremath{{\bold B}}}

\newcommand\bXi{\ensuremath{{\bm \Xi}}}

\newcommand\bF{\ensuremath{{\bold F}}}

\newcommand\bd{\ensuremath{{\bm d}}}
\newcommand\bD{\ensuremath{{\bold D}}}

\newcommand\bu{\ensuremath{{\bm u}}}
\newcommand\bv{\ensuremath{{\bm v}}}

\newcommand\bPhi{\ensuremath{{\bm \Phi}}}

\newcommand\bV{\ensuremath{{\bold V}}}

\newcommand\bs{\ensuremath{{\bm s}}}

\newcommand\bE{\ensuremath{{\bold E}}}

\newcommand{\diag}{\mathrm{diag}}

\newcommand\bQ{\ensuremath{{\bold Q}}}

\newcommand{\bI}{{\bold I}}

\begin{document}

\title{Multi-Resolution Codebook Design and Multiuser Interference Management for Discrete XL-RIS-Aided Near-Field MIMO Systems}

\author{Qian Zhang,~\IEEEmembership{Graduate Student Member,~IEEE,}
		Zheng Dong,~\IEEEmembership{Member,~IEEE,}	
		Yufei Zhao,
		Yao Ge,~\IEEEmembership{Member,~IEEE,}
		Yong Liang Guan,~\IEEEmembership{Senior Member,~IEEE,}
		Ju Liu,~\IEEEmembership{Senior Member,~IEEE,}
		and 
		Chau Yuen,~\IEEEmembership{Fellow,~IEEE}
        % <-this % stops a space
\thanks{
	This work was supported in part by the National Natural Science Foundation of China under Grant 62071275; in part by Shandong Provincial Natural Science Foundation under Grant ZR2023LZH003; in part by the China Scholarship Council; in part by Ministry of Education, Singapore, under its MOE Tier 2 (Award: T2EP50124-0032), in part by A*STAR under the RIE2025 Industry Alignment Fund–Industry Collaboration Projects (IAF-ICP) Funding Initiative (Award: I2501E0045), as well as cash and in-kind contribution from the industry partner(s).
	The corresponding authors: Ju Liu; Zheng Dong. E-mail: \{juliu, zhengdong\}@sdu.edu.cn.}% <-this % stops a space
\thanks{Qian Zhang, Ju Liu, and Zheng Dong are with School of Information Science and Engineering, Shandong University, Qingdao 266237, China (e-mail: qianzhang2021@mail.sdu.edu.cn; juliu@sdu.edu.cn; zhengdong@sdu.edu.cn).}
\thanks{Yufei Zhao, Yong Liang Guan, and Chau Yuen are with School of Electrical and Electronics Engineering, Nanyang Technological University, Singapore 639798 (e-mail: yufei.zhao@ntu.edu.sg; yao.ge@ntu.edu.sg; EYLGuan@ntu.edu.sg; chau.yuen@ntu.edu.sg).}
\thanks{Yao Ge is with the Continental-NTU Corporate Lab, Nanyang Technological University, Singapore 639798 (e-mail: yao.ge@ntu.edu.sg).}
}

\markboth{}%
{Shell \MakeLowercase{\textit{et al.}}: A Sample Article Using IEEEtran.cls for IEEE Journals}


\maketitle

\begin{abstract}
Extremely large-scale reconfigurable intelligent surface (XL-RIS) can effectively overcome severe fading and provide higher communication performance. 
However, current research on XL-RIS overlooks the discrete phase-shift characteristics of RIS in practical systems, which will result in significant performance degradation.
In this paper, we investigate near-field communication schemes assisted by XL-RIS with discrete phase shifts.
Specifically, we propose a hierarchical beam training method to obtain the user channel state information (CSI), and develop the jointly optimized codebook construction (JOCC) method and separately optimized codebook construction (SOCC) method for base station (BS) precoding and XL-RIS phase shifts, respectively. 
With JOCC, the most superior beam training performance can be obtained.
With SOCC, higher performance than the single-antenna BS codebook can be obtained at a similar complexity.
Further, we propose a flexible multiuser interference management (IM) method that is simple to solve. 
The IM method uses adaptive gain matrix approximation to take into account user fairness and can be solved in closed-form iterations. 
In addition, we extend the proposed method to a hybrid precoding design.
Simulation results demonstrate that the proposed multi-resolution codebook construction method can obtain more accurate beam patterns and user CSI, and the proposed IM method obtains superior performance over the benchmark methods.

\end{abstract}

\vspace{-0.1cm}
\begin{IEEEkeywords}
Discrete extremely large-scale reconfigurable intelligent surface (XL-RIS), near-field communication, multi-resolution codebook, hierarchical beam training, multiuser interference management, closed-form solution.
\end{IEEEkeywords}

\section{Introduction}
\IEEEPARstart{R}{econfigurable} intelligent surface (RIS) is a two-dimensional planar array of passive programmable logic units that can be artificially controlled to achieve beamforming in different directions, thereby reconfiguring the wireless transmission environment for superior communication quality~\cite{wu2020risreview,Basar2024RIS,Wu2019IRS,Ahmed2025ActiveRIS}.
RIS is one of the most promising technologies for sixth-generation (6G) communications due to its passive and easy-to-deploy nature.
Based on the significant advances, RIS has attracted widespread attention from both the industry and academia.
At present, RIS has been applied in many fields, such as physical layer security~\cite{zhang2023secrecy}, non-terrestrial communications~\cite{Yuan2023NonTerrestrial}, interference channel degrees-of-freedom (DoF) enhancement~\cite{Zheng2023ARIS}, and integrated sensing and communication~\cite{Yu2024RIS_ISAC}.
However, the multiplicative fading introduced by the RIS prevents significant performance gains from being achieved~\cite{Zhang2023ActiveRIS}.
Fortunately, it has been shown in~\cite{Zhang2023ActiveRIS,Zhang2025BD_RIS} that passive RIS can achieve a square-law signal-to-noise ratio (SNR) increase, and thus fading compensation can be effectively achieved with larger RIS array sizes.
For passive RIS, the larger RIS arrays do not necessarily result in greater hardware losses but can offer more advantages such as channel hardening~\cite{bjornson2021rischannel} as well as higher communication rate gain.
Consequently, extremely large-scale RIS (XL-RIS) is expected to play a crucial role in achieving high-quality communication for 6G networks.

At present, there are some research works related to XL-RIS.
The key distinction between XL-RIS and conventional RIS systems lies in the increased array aperture, which leads to a larger Rayleigh distance and consequently extends the near-field (NF) region~\cite{Yang2024XL_RIS,Yang2023CE_XL_RIS,Gong2024HMIMO}.
Therefore, in XL-RIS systems, users are more likely to fall within the NF communication range, leading to significant changes in the design philosophy of communication systems.
Firstly, within the NF communication range, electromagnetic waves no longer propagate as plane waves but rather as spherical waves~\cite{Gong2024NFMIMO,An2024NFComm}. 
As a result, traditional methods based on plane wave assumptions, such as channel modeling, channel estimation, and precoding design, are no longer applicable~\cite{Cui2023CommunMag,Liu2023NearField}.
Secondly, XL-RIS will lead to a significant increase in the dimensions of the channel matrix, requiring more pilot overhead to obtain accurate channel state information (CSI), which will reduce the available signal transmission resource within a block time slot and result in higher computational overhead.
Therefore, current research on extremely large-scale antenna systems primarily focuses on how to efficiently obtain user CSI using low-complexity and physical resource/time-frequency efficient methods, such as~\cite{Lu2024XL_MIMO,Wei2022Chinacomm,Shen2023Multi_beam,Zhang2022FastBeamTraining,Lv2024codebook}.
In these studies, beam training has become the preferred technique for obtaining user CSI. 

Beam training involves performing beam scanning based on a pre-designed codebook and determining user CSI using the received SNR as feedback. 
When electromagnetic waves arrive as plane waves, the user is in the far-field (FF) communication region, and the channel is only related to the angle of departure (AOD) of the antenna array.
In this case, beam training can be performed using an orthogonal discrete Fourier transform (DFT) codebook to capture the physical angle information of the user's channel paths, thereby obtaining the user CSI~\cite{Ke2020CS,Lim2020EfficientBeamT}.
However, in near-field (NF) systems, the user's channel is not only related to the array's AOD but also to the distance between the user and the array. 
In this case, the beam focuses on a region constrained by both angle and distance, resulting in a grid-like pattern across the entire plane, which will lead to the number of codewords to skyrocket, causing a significant training overhead~\cite{Wei2022XL_MIMO}.
An effective approach for reducing the beam training overhead is the hierarchical beam training method.
In~\cite{Wei2022Chinacomm}, Wei {\it et al.} proposed a codebook design and hierarchical beam training method for XL-RIS-aided NF single-input single-output (SISO) systems.
In~\cite{Shen2023Multi_beam}, Shen {\it et al.} proposed a multi-beam design method, building on the approach from \cite{Wei2022Chinacomm}, to serve multiple users at different locations simultaneously.
In~\cite{Lv2024codebook}, Lv {\it et al.} considered an XL-RIS-aided single-user multiple-input multiple-output (MIMO) communication scenario with both NF and mixed-field communication. 
In this scenario, the RIS phase shifts are optimized through a hierarchical beam training process, while the BS precoding vector is solved by using the weighted minimum-mean-square-error (WMMSE) algorithm~\cite{Shi2011WMMSE}.
However, although these studies considered hierarchical beam training strategies, none of them designed multi-resolution codebooks based on this concept. 
Instead, the codebook designs are set by selecting single sampling points within the regions divided at each level.
In fact, to ensure the accuracy of hierarchical beam training, different beam widths should be used in beam training at different levels~\cite{Xiao2016HCodebook_Criteria,Lu2024XL_MIMO}. 
Furthermore, existing studies primarily focus on codebook design and beam training for continuous phase-shift XL-RIS.

In practice, the phase shifts of RIS are not continuous but are discrete, determined by the constraints of practical hardware design. 
Therefore, codebooks designed based on continuous phase-shift RIS models will result in beam pattern mismatches in practical RIS systems, leading to significant errors in the acquired user CSI. 
In such cases, the precoding cannot be effectively designed, which in turn causes severe signal interference among multiple users, leading to significant performance degradation.
To the best of the authors' knowledge, there are currently no codebook design methods that consider practical RIS models with discrete phase shifts in XL-RIS-aided NF systems.
Additionally, while \cite{Shen2023Multi_beam} proposed a multi-beam design method to achieve multiuser communication,  inter-user interference is not effectively managed/minimized. 
In other words, this method may result in significant inter-user interference in non-orthogonal multiple-access communication, leading to a sharp decline in performance.
Currently, in MIMO systems, the multiuser sum rate maximization method is commonly used to improve spectral efficiency and reduce inter-user interference~\cite{Zhang2024PracticalRIS,Guo2020ProxLinear}, which is an effective approach for managing multiuser signal interference. 
However, this method is highly sensitive to users' channel quality. 
Specifically, to achieve the most superior performance, 
all resources will be fully allocated to the user with the best channel quality, resulting in the remaining users being blocked.
In order to ensure fair communication between users, the max-min signal-to-interference-plus-noise ratio (SINR) problem has been proposed and widely studied~\cite{Zhang2025BD_RIS,Nadeem2022MaxMinSINR}.
However, the max-min SINR problem seeks optimal fairness and will allocate more resources to users with poor channel quality thus leading to a decrease in achievable rate.
Furthermore, the sum rate maximization problem has a non-convex objective function and the max-min SINR problem has an implicit non-convex objective function, both of which are difficult non-convex problems to solve.
Currently, common approaches involve using the fractional programming (FP) algorithm \cite{Shen2018FP} or the WMMSE algorithm~\cite{Shi2011WMMSE} to transform it into a convex function for solving, or relaxing the objective function and solving it by using the semi-definite relaxation (SDR) algorithm~\cite{Luo2023SDR}.
Therefore, solving these problems inevitably involves transformations or relaxations of the non-convex problem, which will inevitably lead to a performance loss.

To tackle these challenges, we propose two effective multi-resolution codebook design methods and a flexible multiuser interference management (IM) strategy for discrete XL-RIS-aided NF systems. 
It is worth noting that the proposed method is not only valid for XL-RIS-aided full-digital BS precoding systems but can also be directly extended to the design of XL-RIS-aided BS hybrid precoding systems.
Specifically, our research and main contributions are as follows. \vspace{-0.15cm}
\begin{itemize}
	\item[$\bullet$] Contrary to previous codebook construct methods in XL-RIS-aided single-antenna BS (SA-BS) systems, we propose two multi-resolution codebook construction methods to match the beamwidth of different levels of beam training under discrete XL-RIS-aided multi-antenna BS (MA-BS) systems.  
	Specifically, we propose a jointly optimized codebook construction (JOCC) method and a separately optimized codebook construction (SOCC) method for BS precoding and RIS phase shifts, respectively.
	The JOCC method can achieve one-by-one matching between the BS precoding and the RIS phase shift in the codebook to ensure the best beam training performance, and the SOCC method optimizes the RIS phase shift by presetting the BS precoding to ensure superior performance.
	The complexity and storage requirements of the SOCC method are similar to those of the SA-BS codebook construction method, but its performance is far superior.

	\item[$\bullet$] We propose an effective multiuser IM method that is more flexible and simpler to handle than benchmark methods and provides more intuitive IM.
	The proposed IM method employs desired gain matrix approximation to ensure that the desired user is covered by the high-gain beam while suppressing beam energy leakage to other users.
	The desired gain matrix adaptive update method is proposed to enhance user communication fairness.
	In addition, the multiuser sum rate maximization and the max-min SINR methods, as benchmark schemes, are studied to demonstrate the flexibility and effectiveness of the proposed IM method.

	\item[$\bullet$] We propose an efficient alternating optimization (AO) algorithm to solve the multi-resolution codebook construct problem and the multiuser IM problem.
	For the implicit variable subproblem of optimizing BS precoding, we rephrase it as an equivalent problem with explicit variables and iteratively solve it in a closed form.
	For the phase shift optimization subproblem, we introduce auxiliary variables to decouple complicated discrete phase shift constraints and solve the closed-form solutions of decoupled subproblems based on the increasing penalty dual decomposition (IPDD) algorithm.
	In addition, we extend the proposed method to hybrid precoding systems.

	\item[$\bullet$] We provide detailed simulation results to demonstrate the effectiveness and superiority of the proposed methods. 
	The simulation results show that the hierarchical beam training implemented based on the proposed multi-resolution codebook design method can achieve higher achievable rates and more accurate CSI acquisition than the benchmark methods under existing SA-BS systems. 
	In addition, the proposed IM method can achieve high-quality communication with user fairness based on its simple-to-solve and flexible characteristics, and its advantages are obvious compared to benchmark methods. 
	The hybrid precoding also approximates the full-digital precoding well and achieves accurate beam pattern matching.

\end{itemize}

{\it{Notation:}} 
$||\cdot||_2$ denotes the Euclidean norm, 
$|\cdot|$ denotes the absolute value,
and $\angle(A)$ denotes the angle of $A$.
$\bx \sim {\cal{CN}}\left( \bm 0, \bm\Gamma \right)$ represents that $\bx$ follows the circularly symmetric complex Gaussian (CSCG) distribution, where the mean vector is $\bm 0$ and the covariance matrix is $\bm\Gamma$.
$(\cdot)^{*}$, $(\cdot)^{\rm T}$, $(\cdot)^{\rm H}$, and $(\cdot)^{-1}$ denote the conjugate, transpose, and conjugate transpose operation, matrix inverse operation, respectively.
$\mathbb{C}^{M \times N}$ represents $M \times N$ complex matrix space.
$\odot$ and $\otimes$ denote the Hadamard product and the Kronecker product, respectively.
$\bold{I}_K$ denotes a $K \times K$ identity matrix and $\bm 1_K$ denotes a $K \times 1$ dimensional vector where all elements are 1.
${\rm diag}(\cdot)$ denotes the diagonalization operation and ${\rm vec}(\cdot)$ denotes the matrix vectorization operation.
${\rm Re}\{\cdot\}$ and ${\rm Im}\{\cdot\}$ denote taking the real and imaginary parts of a complex number, respectively.
${\rm unvec}_{N,M}(\cdot)$ denotes the conversion operation of $NM\times 1$ dimensional vector into an $N\times M$ dimensional matrix.
$\Pi_{\cal C}$ denotes the projection from a point to the set ${\cal C}$.
$\cap$ is to take the intersection operation.
$\lfloor \cdot \rfloor$ denotes a downward rounding operation.
${\rm mod}(\cdot,\cdot)$ denotes a remainder operation.

\vspace{-0.3cm}
\section{System Model}
\begin{figure}[t]	
	\centering \includegraphics[width=0.75\linewidth]{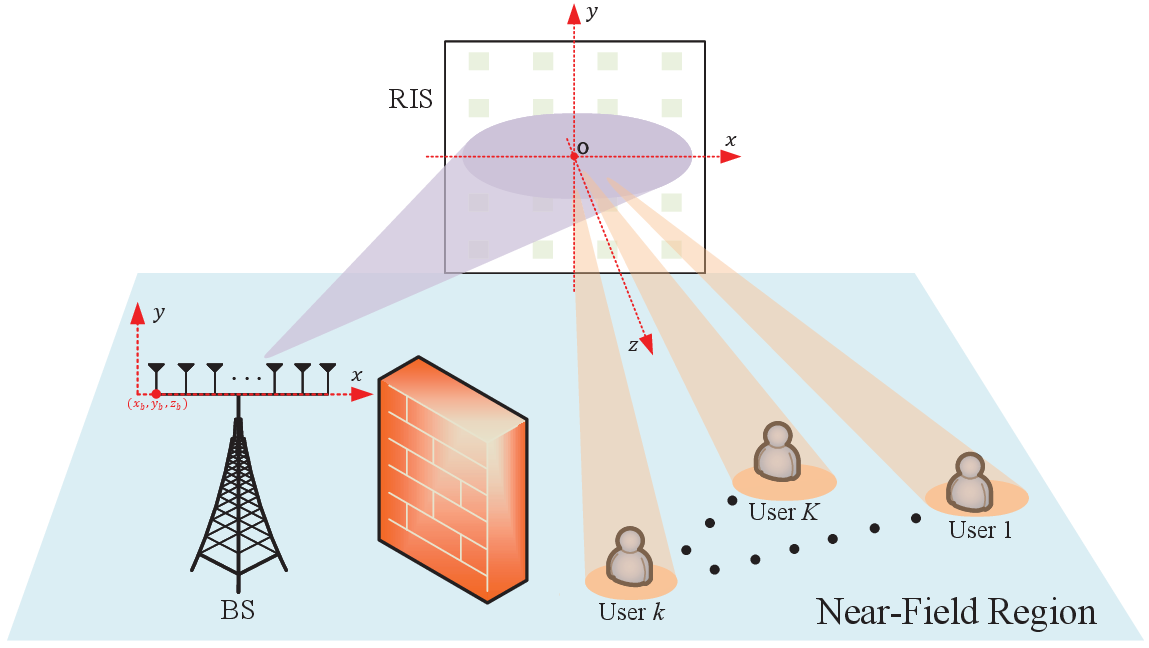}
	\vspace{-0.2cm}
	\caption{The model of XL-RIS-aided multiuser NF MIMO systems.}
	\vspace{-0.6cm}
	\label{fig:System_Model_01}
\end{figure}

As shown in Fig.~\ref{fig:System_Model_01}, we consider a downlink XL-RIS-aided communication system, where the base station (BS) equipped with $M$ antennas serves $K$ single-antenna users assisted by a discrete phase-shift XL-RIS with $N = N_2\times N_1$ elements, and the direct link is blocked.
In the XL-RIS system, the increase in RIS array size will result in a dramatic increase in NF communication range.
The NF and FF boundaries are divided by the Rayleigh distance $\frac{2D_{\rm ris}^2}{\lambda_c}$, where $D_{\rm ris}$ and $\lambda_c$ denote the RIS array aperture and wavelength, respectively.
When $D_{\rm ris}$ is 1.5m and $\lambda_c$ is 0.03m, the Rayleigh distance is 150 meters, then all users within 150 meters are in the NF range.
In addition, according to \cite{Cui2023CommunMag}, the Rayleigh distance $\frac{2D_{\rm ris}^2}{\lambda_c}$ is larger than $\frac{r_{\rm br}r_{\rm ru}}{r_{\rm br} + r_{\rm ru}} $ causing both the BS and the user to be in the NF range, where $r_{\rm br}$ and $r_{\rm ru}$ denote the distance from the BS to the RIS and from the RIS to the user, respectively.
Therefore, we assume the user communication is in the NF communication region.
In addition, we consider a discrete XL-RIS model to match the practical RIS system, which has not yet been studied in XL-RIS-aided NF communications.
Specifically, the phase shift of the XL-RIS is discrete, i.e., the $v$-bit phase shift is limited in the set $\left\{ 0, \frac{2\pi}{2^v},\dots,\frac{(2^v-1)2\pi}{2^v} \right\}$.

\vspace{-0.35cm}
\subsection{Signal Transmission Model}
In the XL-RIS-aided multiuser system, the transmitted signal at the BS is given by 
\begin{equation*}
	\begin{split}
		\bx = \sum_{k=1}^{K} \bw_k s_k,
	\end{split}
\end{equation*}
where $\bw_k$ denotes the precoding vector with respect to user $k$, $\sum_{k=1}^{K} \| \bw_k \|_2^2 \leq P_{\rm max}$, $P_{\rm max}$ denotes the maximum transmission power, and $\bs = [s_1,s_2,\dots,s_K]^{\rm T} \sim {\cal CN}(\bm 0, \bI_K)$ denotes the transmitted data symbol for users.

Then, the received signal of the user $k$ is given by
 \begin{equation*}
 	\begin{split}
 		y_k =& \underbrace{ \bm\phi^{\rm H} \bH_k \bG \bw_k s_k }_{ {\rm{desired}}\, {\rm{signal}} }  + \underbrace{ \sum_{i=1,i\neq k}^{K} \bm\phi^{\rm H} \bH_k \bG \bw_i s_i }_{  {\rm{interference}} \, {\rm{from}} \, {\rm{other}} \, {\rm{users}} }  + \underbrace{n_k}_{ {\rm{noise}} },
 	\end{split}
 \end{equation*}
where $\bH_k = \diag(\bh_k^{\rm H})$, $\bm\phi = \left[ \phi_1,\phi_2, \dots, \phi_N \right]^{\rm T} $ denotes the phase-shift vector of the XL-RIS, $n_k \sim {\cal CN}(0,\sigma_k^2)$ denotes the noise at the user $k$, $\bh_k\in \mathbb{C}^{N\times 1}$ denotes the channel from the RIS to the user $k$, and $\bG \in \mathbb{C}^{N\times M}$ denotes the channel from the BS to the RIS.
Then, the achievable rate of the user $k$ is given by
\begin{equation}
	\begin{split}
		R_k = {\rm log}_2 \left(1 + \frac{| \bm\phi^{\rm H} \bH_k \bw_k |^2}{\sum_{i=1,i\neq k}^{K} | \bm\phi^{\rm H} \bH_k \bw_i |^2 + \sigma_k^2  } \right).
	\end{split}
\end{equation}

Note that to develop effective precoding vectors $\bw_k,k=1,\dots,K$ and phase-shift vector $\bm\phi$, it is necessary to obtain accurate CSI of all users.
Unfortunately, the channel estimation cannot be performed effectively by most of the conventional methods due to the extremely large number of RIS reflecting elements.
Therefore, we aim to acquire user CSI by the beam training method.
It is worth noting that the BS and XL-RIS are usually located on high buildings, and in general, we consider higher carrier frequencies, e.g., 10 GHz, thus there is less scattering in the propagation environment.
Without loss of generality, we only need to determine the physical coordinate of the main path by beam training instead of explicitly estimating the entire channel.
Thus, in the following sections, only the main path is considered, and the corresponding beam training method will be proposed to search for the optimal directional beam to align with the main path.

\vspace{-0.4cm}
\subsection{Near-Field Channel Model}
\begin{figure}[t]	
	\centering \includegraphics[width=0.75\linewidth]{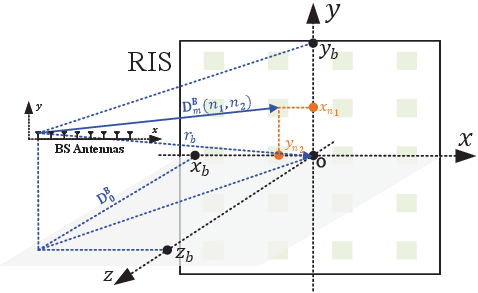}
	\vspace{-0.25cm}
	\caption{Coordinate system model between the BS and the RIS.}
	\vspace{-0.4cm}
	\label{fig:System_Model_BS_RIS}
\end{figure}

\begin{figure}[t]	
	\centering \includegraphics[width=0.9\linewidth]{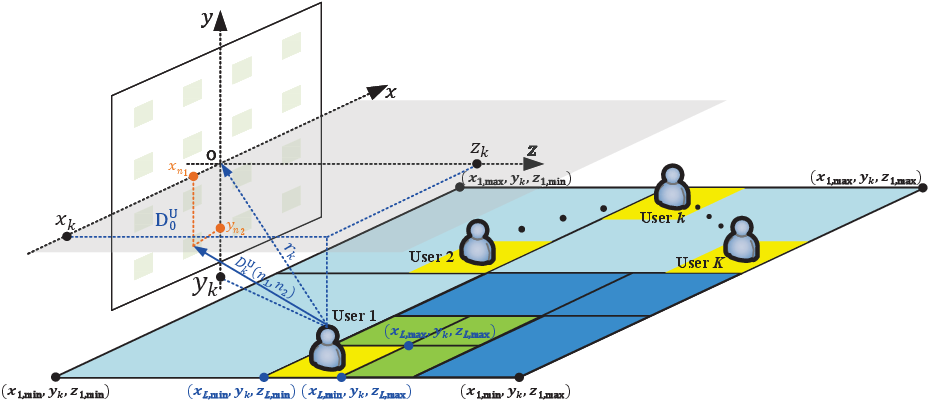}
	\vspace{-0.25cm}
	\caption{Coordinate system model between the RIS and users.}
	\vspace{-0.5cm}
	\label{fig:RIS_User_hierarchical_beam_training}
\end{figure}

\vspace{-0.1cm}
In this subsection, we establish a three-dimensional (3D) Cartesian coordinate system with the center point of the XL-RIS array as the origin, where the RIS is placed in the $x$-$o$-$y$ plane, as shown in Fig.~\ref{fig:System_Model_01}.
In the 3D Cartesian coordinate system shown in Fig.~\ref{fig:System_Model_BS_RIS}, the coordinate position of BS is $(x_b,y_b,z_b)$.
As shown in Fig.~\ref{fig:RIS_User_hierarchical_beam_training}, the coordinate position of the user $k$ is $(x_k,y_k,z_k)$ in the Cartesian coordinate system.
The coordinates of the RIS element are represented as $(x_{n_1}, y_{n_2},0)$ with $x_{n_1}= (n_1 - \frac{N_1+1}{2})d$ and $y_{n_2} = (n_2 - \frac{N_2+1}{2})d$, where $d=\lambda_c/2$, $n_1 = 1,\dots,N_1$, and $n_2 = 1,\dots,N_2$.

The coordinate position of the $m$-th antenna at the BS is $(x_b+(m-1)d, y_b, z_b)$, where $m=1,\dots,M$.
Based on the geometry given in Fig.~\ref{fig:System_Model_BS_RIS}, the NF channel of the main path between the BS and the RIS is modeled as
\begin{equation}
	\begin{split}
		\bG = \left[ \bg_1,\bg_2,\dots,\bg_{M} \right],
	\end{split}
\end{equation}
where $\bg_m \in \mathbb{C}^{N \times 1}$ denotes the channel between $m$-th antenna at the BS and the RIS array;
\begin{equation*}
	\begin{split}
		\bg_m = & \left[\kappa_{m}^{\rm B}(1,1),\dots, \kappa_{m}^{\rm B}(N_1,N_2) \right]^{\rm T} \odot\\
		& \left[e^{-j \frac{2\pi }{\lambda_c}D_{m}^{\rm B}(1,1) }, \dots, e^{-j \frac{2\pi }{\lambda_c}D_{m}^{\rm B}(N_1,N_2) }
		\right]^{\rm T},
	\end{split}
\end{equation*}
where $\kappa_{m}^{\rm B}(n_1,n_2) =  \frac{D_0^{\rm B}}{D_{m}^{\rm B}(n_1,n_2)}$ denotes the free-space large-scale path loss between the $m$-th BS antenna and the $(n_1,n_2)$-th RIS element \cite{Han2020NF_Channel}
and $D_0^{\rm B}$ denotes the distance from the BS to the RIS plane, i.e., $D_0^{\rm B} = |z_b|$ and
$ D_m^{\rm B}(n_1,n_2) = \sqrt{ \left(x_b+\frac{(2m + N_1 - 2n_1 - 1)d}{2} \right)^2 + \left(y_b + \frac{(N_2 - 2n_2 + 1)d}{2} \right)^2 + z_b^2 }$.

Based on the geometry given in Fig.~\ref{fig:RIS_User_hierarchical_beam_training}, the NF channel between the RIS array and user $k$ is modeled as
\begin{equation*}
	\begin{split}
		 \bh_k(x_k,y_k,z_k) =& \left[\kappa_k^{\rm U}(1,1),\dots, 
		\kappa_k^{\rm U}(N_1,N_2)
		\right]^{\rm T} \odot  \\
		& \left[ e^{-j \frac{2\pi }{\lambda_c}D_k^{\rm U}(1,1) },\dots,
		e^{-j \frac{2\pi }{\lambda_c}D_k^{\rm U}(N_1,N_2) }
		\right]^{\rm T},
	\end{split}
\end{equation*}
where $\kappa_k^{\rm U}(n_1,n_2) =  \frac{D_0^{\rm U}}{D_k^{\rm U}(n_1,n_2)}$ denotes the free-space large-scale path loss between the $(n_1,n_2)$-th RIS element and user $k$ \cite{Han2020NF_Channel}
and $D_0^{\rm U}$ denotes the distance from user $k$ to the RIS plane, i.e., $D_0^{\rm U} = |z_k|$ and
$D_k^{\rm U}(n_1,n_2) = \sqrt{ \left(x_k + (\frac{N_1+1}{2} - n_1)d\right)^2 + \left( y_k + (\frac{N_2 + 1}{2} - n_2)d \right)^2 + z_k^2 }$.

\section{Hierarchical Beam Training And Multi-Resolution Codebook Design}
\subsection{Hierarchical Beam Training}
\begin{figure}[t]	
	\centering 
	\includegraphics[width=0.9\linewidth]{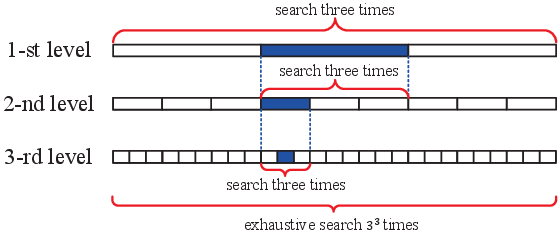}
	\vspace{-0.3cm}
	\caption{Example of the beam training when $L = 3$ and $S = 3$.}
	\vspace{-0.3cm}
	\label{fig:Hierarchical_beam_training}
\end{figure}

\begin{figure}[t]	
	\vspace{-0.25cm}
	\centering 
	\includegraphics[width=0.7\linewidth]{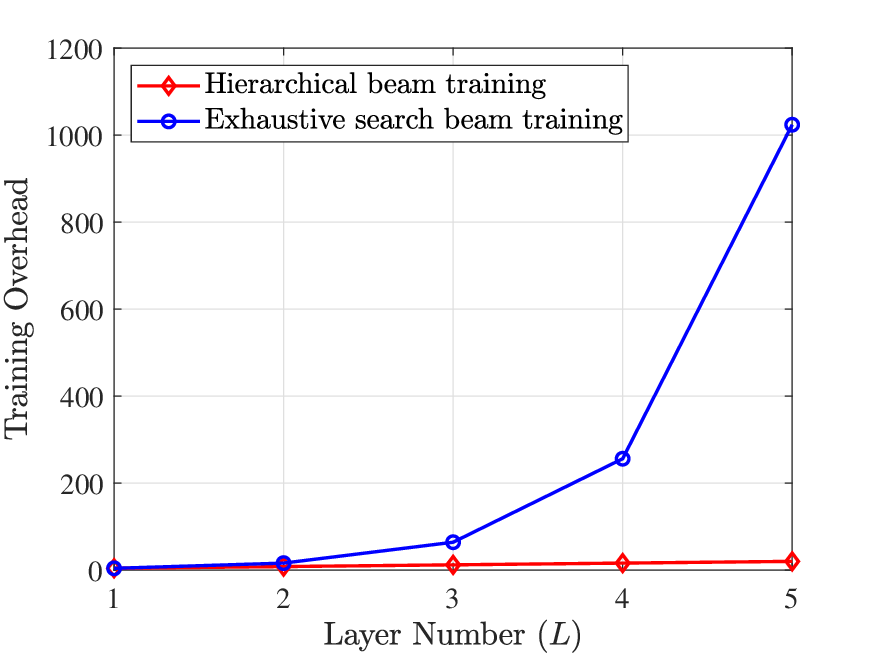}
	\vspace{-0.25cm}
	\caption{Training overhead under different beam training methods, $S=4$.}
	\vspace{-0.6cm}
	\label{fig:Training_overhead}
\end{figure}

The NF channel response implies that the beam training achieved by the optimal codeword should rely on the positions of the BS antenna and users. 
In order to reduce the complexity of beam training, we use a hierarchical beam training method to obtain the main path information of the cascaded channel, as shown in Fig.~\ref{fig:RIS_User_hierarchical_beam_training}.
We assume that there is a total of $L$ level of search, and $S$ regions are searched in each level.
Then, using the hierarchical beam training method, the training overhead for searching the user coordinate region is $S\times L$.
However, using the conventional exhaustive search beam training method, the training overhead for searching the user coordinate region is $S^L$.
In order to better understand the overhead of hierarchical beam training and exhaustive search beam training, we take the example of a 3-level search with 3 regions at each level, as shown in Fig.~\ref{fig:Hierarchical_beam_training}.
It can be seen that three regions are searched at each level in the process from the first level to the third level, so a total of $3\times 3$ searches are required for hierarchical beam training.
When using exhaustive search beam training, all regions of the third level need to be searched, thus a total of $3^3$ times searches are required.
Therefore, the proposed hierarchical beam training method roughly decreases $S^{L-1}/L$ times overhead compared to the exhaustive search beam training, as shown in Fig.~\ref{fig:Training_overhead}.
However, previous FF-based multi-resolution codebooks are not applicable to NF beam training as they will lead to beam mismatch. 
In addition, due to the different beam widths for different beam training levels, it is also crucial to design a multi-resolution codebook suitable for hierarchical beam training.
Therefore, we design multi-resolution codebooks to adapt to different levels of beam training.


\subsection{Multi-Resolution Codebook Design}
\begin{figure}[t]	
	\centering 
	\includegraphics[width=0.8\linewidth]{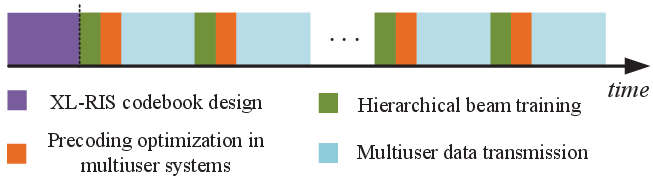}
	\vspace{-0.35cm}
	\caption{The proposed whole communication protocol for codebook design, hierarchical beam training, precoding optimization, and data transmission.}
	\label{fig:Communication_procedure}
	\vspace{-0.6cm}
\end{figure}

In this subsection, we design different resolution codebooks to match the beam training of different levels based on the range of distances sampled between the levels.
The basic idea of our proposed codeword design is to achieve as high a beam gain as possible in the target region whilst suppressing the beam gain in the non-target region.
Thus, the codeword can be obtained by approximating the designed beam pattern to the desired counterpart.
It is worth noting that the precoding of the MA-BS in the codebook design has an impact on beam training. 
In order to avoid the effect of multiple antennas at the BS, the current studies mainly consider the equivalent SA-BS to realize the codebook design of the system, e.g., \cite{Wei2022Chinacomm,Shen2023Multi_beam}.
However, the precoding design of MA-BS is crucial in NF communications, especially high-frequency communications.
Then how to jointly design the BS precoding and RIS phase shifts to construct a multi-resolution codebook is a critical issue. 
Therefore, we propose two multi-resolution codebook construction methods in the XL-RIS-aided MA-BS system to match hierarchical beam training, {\it 1) jointly optimized codebook construction (JOCC)} and {\it 2) separately optimized codebook construction (SOCC)}.
The constructed codebook can be {\it pre-configured} in Micro-Controller Units (MCUs) and then the millisecond RIS beam pattern configuration can be realized through {\it parallel processing}, thus ensuring efficient beam training, as detailed in our previous work~\cite{Zhao2024RIS_Demo}.
Then, after obtaining the user CSI through beam training, we can implement the joint optimization of BS precoding and RIS phase shift to achieve effective multiuser communication.
Fig.~\ref{fig:Communication_procedure} shows the entire communication protocol.

\subsubsection{Jointly Optimized Codebook Construction (JOCC)}
Generally, the BS and the RIS are deployed with a fixed location, thus the relative positions of the two are known.
In addition, without loss of generality, we consider all users at the same height, i.e., the users are located in the $x$-$o$-$z$ plane with fixed $y_k=y_u,\forall k$.
Therefore, when designing the multi-resolution codebook, we mainly consider the exploration of regions in the $x$-axis direction and $z$-axis direction.
The signal transmitted by the BS for beam training is $\bx_t = \bw_t s $, where $\bw_t\in \mathbb{C}^{M\times 1}$ denotes the precoding vector at the BS and $s$ denotes the data symbol for beam training.
Then, the signal received by the user is given by
\begin{equation}\label{recieved_signal}
	\begin{split}
		y_u = \bm\phi^{\rm H} \bC(x_u,y_u,z_u) \bw_t s + n_u,
	\end{split}
\end{equation}
where $\bC(x_u,y_u,z_u) = \diag\left( \bh_u(x_u,y_u,z_u) \right)^{\rm H} \bG \in \mathbb{C}^{N\times M}$ in which $\bh_u(x_u,y_u,z_u)\in \mathbb{C}^{N\times 1}$ denotes the channel from the RIS to the user's coordinate $(x_u,y_u,z_u)$ and $n_u\sim {\cal CN}\left( 0, \sigma_u^2 \right)$ denotes the noise at the user.

In beam training, we identify the CSI corresponding to the codeword that maximizes $\left| \bm\phi^{\rm H} \bC_(x_i,y_u,z_k) \bw_t \right|^2 \big/ \sigma_u^2$ as the user's CSI estimation.
Therefore, in hierarchical beam training, the high-gain beam range at each level should cover the user position.
Based on this, we represent the cascaded channel of the main path with respect to each sampled point as
\begin{equation}
	\begin{split}
		\bB_1 = \left[ \bC(x_1,y_u,z_1)^{\rm T}, \dots, \bC(x_{S_x},y_{u},z_{S_z})^{\rm T} \right]^{\rm T},
	\end{split}
\end{equation}
where $S_x$ and $S_z$ denote the number of samples along the $x$ and $z$ axis, respectively.

Then, we assume the distance region at the $l$-th level is from $x_{l,{\rm min}}$ to $x_{l,{\rm max}}$ and from $z_{l,{\rm min}}$ to $z_{l,{\rm max}}$.
As shown in Fig.~\ref{fig:RIS_User_hierarchical_beam_training}, the coordinate of every sampled point at $l$-th level is given by
\begin{equation*}
	\begin{split}
		x_{l,s_x} &= x_{l,{\rm min}} + \left(s_x - \frac{1}{2} \right) \frac{x_{l,{\rm max}} - x_{l,{\rm min}}}{S_{l}^x},\, s_x = 1,\dots,S_{l}^x, \\
		z_{l,s_z} &= z_{l,{\rm min}} + \left(s_z- \frac{1}{2} \right)\frac{z_{l,{\rm max}} - z_{l,{\rm min}}}{S_{l}^z},\, s_z = 1,\dots,S_{l}^z,
	\end{split}
\end{equation*}
where $S_l^x$ and $S_l^z$ denote the number of samples along the $x$ and $z$ axes, respectively.

Suppose the distance coverage of codeword are 
\begin{equation}
	\begin{split}
		{\cal D}_{x,l} &\triangleq \left[ x - \Delta x_l/2 , ~x + \Delta x_l/2 \right], \\
		{\cal D}_{z,l} &\triangleq \left[ z - \Delta z_l/2 , ~z + \Delta z_l/2 \right],
	\end{split}
\end{equation}
where $\Delta x_l$ and $\Delta z_l$ are the sampled steps in different axis directions for the $l$-th level. The desired beam pattern is denoted as $\bp\odot \bp^{\nu}$ as follows
\begin{equation} \label{idealBeampattern}
	\begin{split}
		&\bp = \left[ p_1(x,y_u,z),\dots, p_{S_x S_z}(x,y_u,z) \right]^{\rm T}, \\
		&\bp^{\nu} = \left[ p_{1}^{\nu}(x,y_u,z),\dots, p_{S_x S_z}^{\nu}(x,y_u,z) \right]^{\rm T},
	\end{split}
\end{equation}
where $p_{i}(x,y_u,z) $ denotes the desired amplitude gain; $ p_{i}^{\nu}(x,y_u,z) = e^{j \nu_i}$ denotes the desired beam phase, which can be flexibly designed to improve the DoF of the codeword design.
The desired amplitude gain $ p_{i}(x,y_u,z) $ can be further given by
\begin{equation} 
	p_{i}(x,y_u,z)  = 
	\begin{cases}
		C_{g},~x \in {\cal D}_{x,l},~z\in {\cal D}_{z,l}, \\
		0,~~~x \notin {\cal D}_{x,l},~z\notin {\cal D}_{z,l},
	\end{cases}
\end{equation}
where $C_{g} >0 $ is a given beam amplitude gain for the desired beam pattern.
The amplitude gains $ p_{i}(x,y_u,z) $ of the desired beam pattern within the target distance coverage is as high and fixed as possible, while the other beam gains are as zero as possible.
It can be noted that the basic idea of our proposed desired beam pattern for different levels coincides with the hierarchical codebook design criterion mentioned in~\cite{Xiao2016HCodebook_Criteria}.

Based on this, we further formulate the multi-resolution codebook design problem as
\begin{equation}\label{codeword_design_multi_antenna}
	\begin{split}
		&\min_{ \bw_t,\, \bm\phi, \bp^{\nu} }~ \left\| \bm\phi^{\rm H}\widetilde\bA_1(\bw_t)  - \bp^{\rm T} \odot \bp^{\nu,{\rm T}} \right\|_2^2 \\
		\mbox{s.t.}	
		&~ {\cal C}_{w_t}: \left\| \bw_t \right\|_2^2 \leq P_{\rm max}, \\
		&~ {\cal C}_{p}: |p_i^{\nu}| = 1,\, i= 1,\dots,S_xS_z,~~ {\cal C}_{\phi}: \phi_n=e^{j\theta_n} ,\\
		&~ {\cal C}_{\theta}:\theta_n \in \left\{ 0, \frac{2\pi}{2^v},\dots,\frac{(2^v-1)2\pi}{2^v} \right\},\, n = 1,\dots,N,
	\end{split}
\end{equation}
where $\widetilde\bA_1(\bw_t) = {\rm unvec}_{N,S_x S_z}\left( \bB_1 \bw_t \right)$.

\subsubsection{Separately Optimized Codebook Construction (SOCC)}
Before designing the NF separately optimized codebook, we first review the codebook design methodology for the FF scenario of RIS-aided MIMO systems.
In this case, the main path channel between the BS and the RIS can be modeled as $\bG^{\rm F} = \ba_r \ba_b^{\rm H}$, where $\ba_r \in \mathbb{C}^{N\times 1}$ and $\ba_b \in \mathbb{C}^{M\times 1}$ denote the steering vectors of the RIS array and the BS antenna array, respectively.
We can design the BS precoding codebook by directly aligning the BS-RIS channel $\bG^{\rm F}$ as follows
\begin{equation}\label{far_field_alignment}
	\begin{split}
		\max_{ \bw_t }~ \left\| \bG^{\rm F}\bw_t \right\|_2^2~~
		\mbox{s.t.}	
		~ \left\| \bw_t \right\|_2^2 \leq P_{\rm max}.
	\end{split}
\end{equation}
\vspace{-0.5cm}

The above problem can be equivalently transformed into the following problem
\begin{equation}\label{far_field_alignment_2}
	\begin{split}
		\max_{ \bw_t }~ \left| \ba_b^{\rm H} \bw_t \right|^2~~
		\mbox{s.t.}	
		~ \left\| \bw_t \right\|_2^2 \leq P_{\rm max}.
	\end{split}
\end{equation}
\vspace{-0.5cm}

By the Cauchy-Schwarz inequality, we have $\left| \ba_b^{\rm H} \bw_t \right|^2 \leq \left\| \ba_b \right\|_2^2 \left\| \bw_t \right\|_2^2 $ and the equal sign is taken when $\ba_b^{\rm H}$ and $\bw_t$ are in the same direction.
Therefore, the globally optimal solution of the problem~\eqref{far_field_alignment} is given by
\begin{equation}
	\bw_t^{\star} = \frac{\ba_b^{\rm H}}{\| \ba_b \|_2}\sqrt{P_{\rm max}},
\end{equation}
which is fully consistent with the codebook design in the existing works~\cite{Wei2022Chinacomm,Shen2023Multi_beam,Lv2024codebook}.

Similarly, based on this principle, we can optimize the BS precoding and RIS phase shifts separately thus achieving a lower complexity design than the JOCC method.

\noindent 
$\blacklozenge$ SOCC Step-1: We first design the BS precoding aligned BS-RIS channel $\bG$, given by
\begin{equation}\label{near_field_alignment}
	\begin{split}
		\max_{ \bw_t }~ \left\| \bG\bw_t \right\|_2^2~~
		\mbox{s.t.}	
		~ \left\| \bw_t \right\|_2^2 \leq P_{\rm max}.
	\end{split}
\end{equation}
\vspace{-0.5cm}

Problem \eqref{near_field_alignment} is a standard Rayleigh quotient problem.
Therefore, the optimal solution is given by
\begin{equation} \label{optimal_w_t}
	\begin{split}
		\bw_t^{S} = \frac{\bv_{\rm max}(\bG^{\rm H} \bG)}{\|\bv_{\rm max}(\bG^{\rm H} \bG) \|_2}\sqrt{P_{\rm max}} ,
	\end{split}
\end{equation}
where $\bv_{\rm max}(\bG^{\rm H} \bG)$ is the eigenvector corresponding to the largest eigenvalue of the matrix $\bG^{\rm H} \bG$.

\noindent 
$\blacklozenge$ SOCC Step-2: 
Based on~\eqref{optimal_w_t}, we next design the RIS phase shifts, given by
\begin{equation}\label{codeword_design_multi_antenna_SOCC}
	\begin{split}
		\min_{ \bm\phi\in {\cal C}_\phi \cap {\cal C}_\theta, \bp^\nu \in {\cal C}_p }&~ \left\| \bB^{\rm H} \bm\phi  - \bp \odot \bp^{\nu} \right\|_2^2,
	\end{split}
\end{equation}
where $\bB = \left[ \bc(x_1,y_u,z_1),\bc(x_1,y_u,z_2),\dots, \bc(x_{S_x},y_u,z_{S_z}) \right]$ and $\bc(x_i,y_u,z_k) = \diag\left(\bh_u(x_i,y_u,z_k) \right)\bG \bw_t^{S} $.

\subsubsection{Performance: JOCC vs. SOCC  }
From problems~\eqref{codeword_design_multi_antenna} and~\eqref{codeword_design_multi_antenna_SOCC}, we can summarize that both the JOCC problem and the SOCC problem have the following optimization problem form.
\begin{equation} \label{summ_codebook}
	\begin{split}
		\min_{\Lambda}~ \left\| \bm\phi^{\rm H}\widetilde\bA_1(\bw_t)  - \bp^{\rm T} \odot \bp^{\nu,{\rm T}} \right\|_2^2,
	\end{split}
\end{equation}
where $\Lambda = \left\{ \bm\phi, \bw_t, \bp^\nu \right\}$ for the JOCC problem and $\Lambda = \left\{ \bm\phi, \bp^\nu \right\}, \bw_t = \bw_t^S$ for the SOCC problem.

We consider the optimal value of the JOCC problem and the SOCC problem as $f_J^{\star}$ and $f_S^{\star}$, respectively.
Assume that $\left\{\bm\phi^S, (\bp^{\nu})^S  \right\}$ is optimal solution to the SOCC problem.
Then, we can obtain that $\left\{\bm\phi^S, (\bp^{\nu})^S, \bw_t^S  \right\}$ is a feasible solution to the JOCC problem.
When $\left\{\bm\phi^S, (\bp^{\nu})^S, \bw_t^S  \right\}$ is also an optimal solution to the JOCC problem, we have $f_J^{\star} = f_S^{\star}$.
When $\left\{\bm\phi^S, (\bp^{\nu})^S, \bw_t^S  \right\}$ is not an optimal solution to the JOCC problem, then there must exist another feasible solution to the JOCC problem that can make the objective function value lower.
Therefore, we can obtain
\begin{equation}
	\begin{split}
		f_J^{\star} \leq \left\| (\bm\phi^S)^{\rm H}\widetilde\bA_1(\bw_t^S)  - \bp^{\rm T} \odot (\bp^\nu)^{S,{\rm T}} \right\|_2^2 = f_S^{\star}.
	\end{split}
\end{equation}
\vspace{-0.4cm}

{\it Remark 1: }
The performance of the SOCC method depends on the magnitude of the singular values of the channel $\bG$. The more dominant components are occupied by the largest singular values of $\bG$, then the performance of the SOCC method is closer to that of the JOCC method. 
In addition, the codebook construction in existing XL-RIS-aided NF communication systems is mainly considered under a SA-BS system~\cite{Wei2022Chinacomm, Shen2023Multi_beam}.
The proposed SOCC method can achieve far better performance than it can with a complexity close to the codebook construction of an SA-BS system, especially in high-frequency communications, where signals are subjected to much more severe fading such that severe degradation of signal quality in SA-BS systems.

{\it Remark 2:} 
In this paper, we aim to present an idea/framework for constructing a joint codebook of the BS precoding and XL-RIS phase shifts, which can be extended to more general scenarios.
For example, if there is a BS-user weak direct link, we can reformulate \eqref{recieved_signal} as $y_u = \widetilde{\bm\phi}^{\rm H} \widetilde\bC \bw_t s + n_u$, where $\widetilde{\bm\phi} = [\bm\phi^{\rm T},1]^{\rm T}$, $\widetilde\bC = [\bC^{\rm H},\bm f_u]^{\rm H}$, and $\bm f_u$ denotes the BS-user channel.
Then, we can modify the variables and channel models in problems~\eqref{codeword_design_multi_antenna} and \eqref{codeword_design_multi_antenna_SOCC} to $\widetilde{\bm\phi}$ and $\widetilde\bC$ to construct codebook.
In fact, we can also achieve signal high-gain transmission by precoding only in the BS-RIS channel, which can be approximated as the weak direct link being blocked, thus avoiding some complications contributing to the beam training accuracy.
As another example, if there are multiple antennas at the user, we can reformulate \eqref{recieved_signal} as $y_u = \bu_u^{\rm H} \bH_u^{\rm H} \bPhi \bG \bw_t s + n_u$, where $\bu_u \in \mathbb{C}^{M_U\times 1}$ and $\bH_u \in \mathbb{C}^{N\times M_U}$ denote the combined vector and the channel from RIS to user, respectively, and $M_U$ denotes the number of user antennas. 
The codebook of $\bu_u$, $\bPhi$, and $\bw_t$ can also be jointly constructed by proposed methods, but this inevitably increases the computational and storage requirements.
Another approach is that the user receives the signal and processes it using selective combination or equal-gain combination methods to reduce complexity.

\vspace{-0.5cm}
\section{Proposed Codebook Design Algorithm }
Problem \eqref{codeword_design_multi_antenna} is a non-convex optimization problem with highly coupled variables and constant-modulus discrete constraints, which is very intractable and hard to solve.
To handle it, we first employ the alternating optimization (AO) algorithm to decouple variables $\bw_t$, $\bm\phi$, and $\bp^\nu$, which follows the updates.
\begin{subequations}\label{AO_step}
	\begin{align}
		\bw_t^{\ell+1} = &~\mathop{\rm argmin}\limits_{\bw_t\in {\cal C}_{w_t}} \left\| \bm\phi^{\ell,{\rm H}}\widetilde\bA_1(\bw_t)  - (\bp^{\rm T} \odot \bp^{\nu,{\rm T}})^{\ell} \right\|_2^2, \label{optimize_w_AO}
		\\
		\bm\phi^{\ell+1} = &~\mathop{\rm argmin}\limits_{\bm\phi \in {\cal C}_{\phi}\cap {\cal C}_{\theta} } \left\| \bm\phi^{\rm H}\widetilde\bA_1(\bw_t^{\ell+1})  - (\bp^{\ell} \odot \bp^{\nu,\ell})^{\rm T} \right\|_2^2, \label{optimize_phi_AO}
		\\
		\bp^{\nu,\ell+1} = &~\mathop{\rm argmin}\limits_{\bp^\nu \in {\cal C}_{p} } \left\| \bm\phi^{\ell+1,{\rm H}} \widetilde\bA_1(\bw_t^{\ell+1})  - ( \bp \odot \bp^{\nu} )^{\rm T} \right\|_2^2. \label{optimize_p_AO}
	\end{align}
\end{subequations}

Then, we propose corresponding efficient algorithms to solve these subproblems~\eqref{optimize_w_AO}-\eqref{optimize_p_AO}.
In addition, both the problem \eqref{codeword_design_multi_antenna_SOCC} and the SA-BS codebook construction problem can also be solved by the following problem.

\vspace{-0.3cm}
\subsection{Algorithm for~\eqref{optimize_w_AO}}
The objective function of the problem~\eqref{optimize_w_AO} is implicitly expressed with respect to $\bw_t$ without direct the optimal solution for $\bw_t$.
To deal with it, we construct a function for which $\bw_t$ is expressed explicitly such that it is exactly equal to the objective function of the problem~\eqref{optimize_w_AO}, as follows
\begin{equation} \label{equal_con}
	\begin{split}
		\bm\phi^{\rm H} \widetilde\bA_1\left(  \bw_t \right)  
		=\bw_t^{\rm T} {\rm unvec}_{M,S_xS_z}\left( \bB_2^{\rm T} \bm\phi^* \right),
	\end{split}
\end{equation}
where $\bB_2 = \left[ \bC(x_1,y_u,z_1), \dots, \bC(x_{S_x},y_{u},z_{S_z}) \right]$.

By~\eqref{equal_con}, the problem~\eqref{optimize_w_AO} can be equivalently transformed into the following problem 
\begin{equation} \label{optimize_w_t}
	\begin{split}
		\min_{ \bw_t }~ \left\| \widetilde\bA_2 \bw_t - \bp \odot \bp^{\nu} \right\|_2^2~~{\rm s.t.}~\left\| \bw_t \right\|_2^2 \leq P_{\rm max},
	\end{split}
\end{equation}
where $\widetilde\bA_2 = \left( {\rm unvec}_{M,S_xS_z}\left( \bB_2^{\rm T} \bm\phi^* \right) \right)^{\rm T}$.

Then, by the Lagrangian method, the optimal solution of the problem~\eqref{optimize_w_t} can be derived as
\begin{equation} 
	\bw_t^{\star}  = \left( \widetilde\bA_2^{\rm H} \widetilde\bA_2 + \lambda_w \bI_M \right)^{-1} \widetilde\bA_2^{\rm H} ( \bp\odot \bp^{\nu} ),
\end{equation}
where $\lambda_w \geq 0$ is a Lagrangian multiplier, which can be found by the one-dimensional line search.

\vspace{-0.3cm}
\subsection{Algorithm for~\eqref{optimize_phi_AO}}
The problem~\eqref{optimize_phi_AO} is also a non-convex problem due to the constant-modulus discrete constraints.
To deal with it, we propose a flexible variable separation method to decouple complicated constraints from the complicated objective function, and then the decoupled problem can be solved based on the framework of the IPDD algorithm~\cite{Shi2020PDD}.
Then, we introduce a new optimization variable $\bm\zeta = \left[ \zeta_1,\zeta_2,\dots,\zeta_N \right]^{\rm T}$ such that $\bm\zeta = \bm\phi$.
Therefore, the problem~\eqref{optimize_phi_AO} can be reformulated as
\begin{equation}\label{multi_antenna_equivalent}
	\begin{split}
		&\min_{ \bm\phi,\, \bm\zeta }~ {\cal H}(\bm\phi, \bm\zeta) = \left\| \bm\phi^{\rm H} \widetilde\bA_1(\bw_t) - \bp^{\rm T} \odot \bp^{\nu,{\rm T}} \right\|_2^2 \\
		\mbox{s.t.}	
		&~ \bm\zeta = \bm\phi,\,\zeta_n = e^{j \theta_n},\\
		&~\theta_n \in \left\{ 0, \frac{2\pi}{2^v},\dots,\frac{(2^v-1)2\pi}{2^v} \right\},\, n = 1,\dots,N.
	\end{split}
\end{equation}

Next, we reformulate the problem~\eqref{multi_antenna_equivalent} based on the framework of IPDD as
\begin{equation*}\label{augmented_Lagrangian}
	\begin{split}
		\min_{ \bm\phi,\, \bm\zeta }&~ {\cal H}(\bm\phi, \bm\zeta) + \frac{1}{2\eta} \| \bm\zeta - \bm\phi + \eta\bu \|_2^2  \\
		\mbox{s.t.}	
		&~ {\cal C}_\zeta: \zeta_n = e^{j \theta_n},\, n = 1,\dots,N, \\
		&~{\cal C}_\theta: \theta_n \in \left\{ 0, \frac{2\pi}{2^v},\dots,\frac{(2^v-1)2\pi}{2^v} \right\},\, n = 1,\dots,N,
	\end{split}
\end{equation*}
where $\bu$ is a dual variable and $\eta$ is a penalty factor.
One can be solved by the following update rules.
\begin{subequations}\label{IPDD_step}
	\begin{align}
		\bm\phi^{t+1} = &~\mathop{\rm argmin}\limits_{\bm\phi} {\cal H}(\bm\phi, \bm\zeta^{t}) + \frac{1}{2\eta} \| \bm\zeta^{t} - \bm\phi + \eta\bu^{t} \|_2^2, \label{IPDD_step_phi}
		\\
		\bm\zeta^{t+1} = &~\Pi_{{\cal C}_{\zeta} \cap {\cal C}_{\theta} } \left( \bm\phi^{t+1} - \eta\bu^{t} \right),  \label{IPDD_step_zeta} \\
		\bu^{t+1} =&~\bu^{t} + (\bm\zeta^{t+1} - \bm\phi^{t+1}) \big/ \eta, \label{nu_update}
	\end{align}
\end{subequations}
where $\eta$ is initialized with a large value and then updated at each iteration with $\eta = \epsilon_\eta \eta$, where $0<\epsilon_\eta < 1$.
With a constantly changing $\eta$, $\bm\phi^{t+1}$ and $\bm\zeta^{t+1} $ are gradually approximated until the iteration is terminated when $ \left\| \bm\phi^{t+1} - \bm\zeta^{t+1} \right\|_2^2 \leq \epsilon_\phi$, then an excellent solution is guaranteed in the constraint space.
In this paper, we set $\epsilon_\phi$ to be $10^{-4}$.
A more detailed mathematical analysis of this optimization framework can be found in~\cite{Shi2020PDD}.

Next, we propose a corresponding efficient algorithm to solve the above subproblems~\eqref{IPDD_step_phi}-\eqref{nu_update}, where the globally optimal solution of each subproblem can be derived.

\subsubsection{Solve Subproblem~\eqref{IPDD_step_phi}}
The objective function of the subproblem \eqref{IPDD_step_phi} can be transformed into a standard quadratic form.
In addition, we use $\widetilde\bA_1$ instead of $\widetilde\bA_1(\bw_t)$ in the following for representational convenience.
Then, we reformulate the problem \eqref{IPDD_step_phi} as
\begin{equation} \label{solve_phi}
	\begin{split}
		\min\limits_{\bm\phi}~\bm\phi^{\rm H} \widetilde\bB_1 \bm\phi - 2{\rm Re}\{ \tilde\bb_1^{\rm H}  \bm\phi \} + \tilde c,
	\end{split}
\end{equation}
where $\widetilde\bB_1 = \frac{1}{2\eta}\bI_N + \widetilde\bA_1 \widetilde\bA_1^{\rm H} $, $\tilde\bb_1 = \widetilde\bA_1(\bp^{*} \odot \bp^{\nu,*}) + \frac{1}{2\eta}\bm\zeta^{\ell} + \frac{1}{2}\bu^{\ell}$, and $\tilde c =\left\| \bp^{\rm T} \odot \bp^{\nu,{\rm T}} \right\|_2^2 + \frac{1}{2\eta} \left\| (\bm\zeta^{\ell} + \eta\bu^{\ell})\right\|_2^2$.

Problem \eqref{solve_phi} is a convex problem, which has a unique stationary point~\cite{Boyd2004convex}.
Therefore, we can obtain its globally optimal solution by solving the derivative of the objective function with respect to $\bm\phi$ to be zero.
The globally optimal solution of the problem \eqref{solve_phi} is thus given by
\begin{equation}
	\begin{split}
		\bm\phi^{\star} = \widetilde\bB^{-1} \tilde\bb.
	\end{split}
\end{equation}

\subsubsection{ Solve Subproblem~\eqref{IPDD_step_zeta}}
The subproblem \eqref{IPDD_step_zeta} is a projection problem, which can be tackled by solving the following problem
\begin{equation}\label{projection_phi}
	\begin{split}
		\min_{ \bm\zeta \in {\cal C}_{\zeta} \cap {\cal C}_{\theta} }&~ \left\| \bm\zeta - (\bm\phi^{\ell+1} - \eta\bu^{\ell}) \right\|_2^2,
	\end{split}
\end{equation}
\noindent which is a non-convex problem with discrete manifold constraints.
To solve it, we develop {\it Theorem 1} to obtain the optimal closed-form solutions.

{\it	
	\noindent \textbf{Theorem 1 (Constant-modulus discrete phase projection (CMDPP)):} Define ${\cal C}_v$ as the space of sets consisting of $2^v-1$ complex points uniformly distributed on the unit circle in the domain of complex numbers.
	Assuming that ${\cal C}_v \triangleq \left\{ 1, e^{j\frac{2\pi}{2^v}},\dots,e^{j\frac{(2^v-1)2\pi}{2^v}} \right\}$, then the projection of the complex point $\kappa$ onto the set space ${\cal C}_v$ can be obtained by the following closed-form expression.
	\begin{equation} 
		\Pi_{{\cal C}_v}(\kappa)  = 
		\begin{cases}
			e^{j\frac{(z-1)\pi}{2^{v-1}}}, ~\mbox{if}~ \angle(\kappa) < \frac{(2z - 1)\pi}{2^v}, \\
			e^{j\frac{z\pi}{2^{v-1}}}, ~~~~~\mbox{if}~ \angle(\kappa) \geq \frac{(2z - 1)\pi}{2^v},
		\end{cases}
	\end{equation}
	where $z = \lfloor \frac{2^{v-1} \angle(\kappa) + \pi }{\pi} \rfloor$.
}

\noindent{\it \textbf{Proof:}} see Appendix~A. $\hfill\blacksquare$

Then, the optimal closed-form solution $\bm\zeta^{\star}$ of the problem \eqref{projection_phi} can be derived as
\begin{equation} \label{solution_zeta}
	\zeta_n^{\star}  = 
	\begin{cases}
		e^{j\frac{(z_n-1)\pi}{2^{v-1}}}, ~\mbox{if}~ \angle(\phi_n^{\ell+1} - \eta u_n^{\ell}) < \frac{(2z_n - 1)\pi}{2^{v}}, \\
		e^{j\frac{z_n\pi}{2^{v-1}}} , ~~~\mbox{if}~ \angle(\phi_n^{\ell+1} - \eta u_n^{\ell}) \geq \frac{(2z_n - 1)\pi}{2^{v}},
	\end{cases}
\end{equation}
where $z_n = \lfloor \frac{2^{v-1}\angle(\phi_n^{\ell+1} - \eta u_n^{\ell}) + \pi }{\pi} \rfloor$ for $n=1,\dots,N$.

\begin{algorithm}[t!]
	\caption{Proposed AO Algorithm for \eqref{codeword_design_multi_antenna}}
	\begin{algorithmic}[1]
		\STATE {\bf Input:}  {initialize $\bw_t$, $\bm\phi$, $\bm\zeta$, $\bu$, $\eta$; $\ell=1$}
		\REPEAT
		
		\STATE $\bw_t^{\ell+1}  = \left( \widetilde\bA_2^{\rm H} \widetilde\bA_2 + \lambda_w \bI_M \right)^{-1} \widetilde\bA_2^{\rm H} ( \bp\odot \bp^{\nu} )$;
		
		\STATE \textbf{repeat} 
		
		\STATE {$\quad\, \bm\phi^{\star} = \widetilde\bB^{-1} \tilde\bb$;}
		
		\STATE $\quad$ { update $\bm\zeta^{\star}$ according to~\eqref{solution_zeta};}
		
		\STATE $\quad$ { update $\bu^{\star}$ according to~\eqref{nu_update};}
		
		\STATE $\quad$ { $\eta = \epsilon_\eta \eta$;}
		
		\STATE { \textbf{until} $\left\| \bm\phi^{\star} - \bm\zeta^{\star} \right\|_2 \leq 10^{-4}$;}
		
		\STATE {$\bm\phi^{\ell+1} = \bm\phi^{\star}$;}
		
		\STATE {$\bp^{\nu,\ell+1} = e^{j \angle\left(\widetilde\bA_1^{\rm T}(\bm\phi^{*})^{\ell+1}\right)}$;}
		
		\STATE {$\ell= \ell+1$;}
		
		\UNTIL { stopping criterion is satisfied. }
		
	\end{algorithmic}
\end{algorithm}

\subsection{Algorithm for~\eqref{optimize_p_AO}}
The problem \eqref{optimize_p_AO} can be equivalently transformed into the following problem
\begin{equation} \label{optimize_p}
	\begin{split}
		\min_{\bp^{\nu}}\sum_{i=1}^{S_xS_z} \sqrt{p_i} | \delta_i - p_i^{\nu} |^2~
		{\rm s.t.}\, |p_i^{\nu}| = 1,\, i=1,\dots,S_xS_z,
	\end{split}
\end{equation}
where $\delta_i = \left[\frac{\bm\phi^{{\rm H}} \widetilde\bA_1 }{C_g}\right]_i$ for $i=1,\dots,S_xS_z$.
One can be solved by the following lemma.

{\it	
	\noindent \textbf{Lemma 1 (Phase alignment):} The globally optimal solution of the optimization problem $\min\limits_{e\in \{|e|=1\}}|e - \tilde e|^2$ can be obtained when $e = {\rm mod} (\angle(\tilde e), 2\pi )$.
}

\noindent{\it \textbf{Proof:}} see Appendix~B. $\hfill\blacksquare$

Therefore, the optimal closed-form solution of the problem \eqref{optimize_p} is given by $\bp^{\nu,\star} = e^{j \angle( \widetilde\bA_1^{\rm T} \bm\phi^* )}$.

The proposed AO algorithm for solving problem \eqref{codeword_design_multi_antenna} is summarized in {\bf Algorithm 1}.

\vspace{-0.5cm}
\subsection{Computational Complexity}
\vspace{-0.1cm}
The computational complexity of solving subproblem \eqref{optimize_w_AO} is ${\cal O}(M^3 + MS_xS_z + MNS_xS_z)$; the computational complexity of solving subproblem \eqref{optimize_phi_AO} is ${\cal O}(MNS_xS_z + NS_xS_z)$; the computational complexity of solving subproblem \eqref{optimize_p_AO} are all ${\cal O}(NS_xS_z(M+1))$. Therefore, the overall computational complexity in each iteration of the proposed alternating algorithm is ${\cal O}(M^3+(M+N+MN)S_xS_z)$.

\vspace{-0.5cm}
\section{Multiuser Sum-Rate Maximization and Max-Min SINR}
\vspace{-0.2cm}
Based on the above beam training, the CSI of $K$ users can be obtained, which can support the design of precoding vectors at the BS and phase shifts of the RIS for multiuser communications, as shown in Fig.~\ref{fig:Communication_procedure}.
\vspace{-0.6cm}
\subsection{Sum-Rate Maximization}
In this subsection, we consider the multiuser sum-rate maximization problem in XL-RIS systems with discrete phase shifts to obtain as high a multiuser sum rate as possible, which cannot take into account user fairness.
The multiuser sum rate maximization problem is given by \vspace{-0.1cm}
\begin{equation} \label{sum_rate_maximization}
	\begin{split}
		&\max_{ \bw,\, \bm\phi }~ \sum_{k=1}^{K} {\rm log}_2\, \left(1 + \frac{| \bm\phi^{\rm H} \bH_k \bw_k |^2}{\sum_{i=1,i\neq k}^{K} | \bm\phi^{\rm H} \bH_k \bw_i |^2 + \sigma_k^2  } \right) \\
		&~ \mbox{s.t.}	
		{\cal C}_{w}: \sum_{k=1}^{K}\left\| \bw_k \right\|_2^2 \leq P_{\rm max},~ {\cal C}_{\phi}: \phi_n=e^{j\theta_n} ,\\
		&~{\cal C}_{\theta}: \theta_n \in \left\{ 0, \frac{2\pi}{2^v},\dots,\frac{(2^v-1)2\pi}{2^v} \right\},\, n = 1,\dots,N,
	\end{split}
\end{equation}
where $\bH_k = \diag(\bh_k^{\rm H})\bG$ and $\bw = \left[\bw_1^{\rm T},\bw_2^{\rm T},\dots, \bw_K^{\rm T}  \right]^{\rm T}$.

The objective function of the above problem is non-convex, which can be converted into a convex function for the variables $\bw_k$ or $\bm\phi$, respectively, by the WMMSE method.
Therefore, by WMMSE, the objective $\max\limits_{\bw,\, \bm\phi}\sum_{k=1}^{K} R_k$ can be transformed into the following function \vspace{-0.1cm}
\begin{equation*}
	\begin{split}
		\min\limits_{\bw,\, \bm\phi,\, \bv,\, \bm\rho}~ \sum_{k=1}^{K}\left[\bw_k^{\rm H} \bD \bw_k - 2{\rm Re}\left\{ \bd_k^{\rm H}\bw_k \right\} \right],
	\end{split}
\end{equation*}
where $\bD = \sum_{i=1}^{K} \rho_i |v_i|^2 \bH_i^{\rm H} \bm\phi \bm\phi^{\rm H}\bH_i$ and $\bd_k = \rho_k v_k \bH_k^{\rm H} \bm\phi $.
The optimal solution of $\rho_k$ and $v_k$ can be derived as
\[
v_k^{\star} = \frac{\bm\phi^{\rm H} \bH_k \bw_k }{\sum_{k=1}^{K}|\bm\phi^{\rm H} \bH_k^{\rm H} \bw_i|^2 + \sigma_k^2 },~
\rho_k^{\star} = \left(1-v_k^* \bm\phi^{\rm H} \bH_k \bw_k \right)^{-1}.
\]

%\subsubsection{\textcolor{blue}{Optimize $\bw$}}

By the Lagrangian method, the optimal solution of the problem~\eqref{sum_rate_maximization} with respect to $\bw$ can be derived as
\begin{equation} 
	\bw_k^{\star}  = \left( \bD + \lambda_1 \bI_M \right)^{-1} \bd,~k=1,2,\dots K,
\end{equation}
where $\lambda_1 \geq 0$ is a Lagrangian multiplier.

%\subsubsection{\textcolor{blue}{Optimize $\bm\phi$}}

Problem~\eqref{sum_rate_maximization} with respect to $\bm\phi$ can be rewritten as
\begin{equation} \label{optimize_phi}
	\begin{split}
		\min\limits_{\bm\phi}~ \bm\phi^{\rm H} \hat\bD \bm\phi - 2{\rm Re}\{ \hat\bd^{\rm H} \bm\phi \}~~{\rm s.t.}~ \bm\phi \in {\cal C}_{\phi}\cap{\cal C}_{\theta},
	\end{split}
\end{equation}
where $\hat\bD = \sum_{k=1}^{K} \rho_k |u_k|^2 \bH_k (\sum_{i=1}^{K} \bw_i \bw_i^{\rm H} ) \bH_k^{\rm H} $ and $\hat\bd = \sum_{k=1}^{K} \rho_k u_k^* \bH_k \bw_k $.
It can be solved by the IPDD algorithm.

\vspace{-0.3cm}
\subsection{Max-Min SINR}
In this subsection, we focus on the problem of max-min SINR under XL-RIS systems with discrete phase shifts to obtain as fair a user quality as possible, which cannot be reconciled with maximizing the achievable rate.
The max-min SINR problem is given by
\begin{equation} \label{max_min_SINR}
	\begin{split}
		&\max_{ \bw,\, \bm\phi }~ \min_k \frac{| \bm\phi^{\rm H} \bH_k \bw_k |^2}{\sum_{i=1,i\neq k}^{K} | \bm\phi^{\rm H} \bH_k \bw_i |^2 + \sigma_k^2  } \\
		&~ \mbox{s.t.}	
		{\cal C}_{w}: \sum_{k=1}^{K}\left\| \bw_k \right\|_2^2 \leq P_{\rm max},~ {\cal C}_{\phi}: \phi_n=e^{j\theta_n} ,\\
		&~{\cal C}_{\theta}: \theta_n \in \left\{ 0, \frac{2\pi}{2^v},\dots,\frac{(2^v-1)2\pi}{2^v} \right\},\, n = 1,\dots,N.
	\end{split}
\end{equation}

\vspace{-0.2cm}
By {\it Corollary~3} in~\cite{Shen2018FP}, the problem~\eqref{max_min_SINR} can be transformed into the following problem \vspace{-0.2cm}
\begin{equation*}
	\begin{split}
		&\max_{ \bw,\, \bm\phi,\, z }~ z \\
		&~ \mbox{s.t.}	
		\bw \in {\cal C}_{w},~\bm\phi\in {\cal C}_\phi \cap {\cal C}_\theta,\\
		&~		2 {\rm Re}\{v_k^* \bm\phi^{\rm H} \bH_k \bw_k \} - |v_k|^2 ( \sum_{i\neq k}^{K} | \bm\phi^{\rm H} \bH_k \bw_i |^2 + \sigma_k^2 ) \geq z ,
	\end{split}
\end{equation*}
where $v_k = \frac{\bm\phi^{\rm H} \bH_k \bw_k}{\sum_{i\neq k}^{K} | \bm\phi^{\rm H} \bH_k \bw_i |^2 + \sigma_k^2}$.

The above problem with respect to $\bw$ is a convex problem, which can be solved by CVX.
In addition, the above problem with respect to $\bm\phi$ can be solved by the IPDD algorithm and CVX.
The provided algorithm is a commonly used and concise algorithm for solving this problem at present.

{\it Remark 3:}
The multiuser sum-rate maximization problem and the max-min SINR problem are important problems in the designs of communication systems, which have been widely studied~\cite{Zhang2024PracticalRIS,Zhang2025BD_RIS,Nadeem2022MaxMinSINR,Guo2020ProxLinear}.
However, the multiuser sum-rate maximization method in pursuit of extreme rates will lead to the blocking of users with poor channel quality at low SNR. 
The max-min SINR method in pursuit of extreme fairness will lead to a decrease in total rates. 
In addition, the corresponding algorithms for these problems are complicated and may lead to performance degradation due to the use of relaxation methods.

\vspace{-0.1cm}
\section{Proposed Multiuser Interference Management}
In this section, we aim to propose a multiuser IM method that is simpler to solve and avoids the use of relaxation methods while maintaining user fairness.
Specifically, we divide the proposed IM method into two steps, including 1) desired gain matrix approximation and 2) desired gain matrix adaptive adjustment.

\vspace{-0.3cm}
\subsection{Desired Gain Matrix Approximation }
In order to achieve high-quality communication, the precoding vector at the BS and the phase-shift vector at the XL-RIS should be optimized to align with the high-gain beam to the main path of the desired user.
We assume that the desired beam gain of user $k$ is denoted as $q_k$.
Therefore, the precoding and phase shift optimization problem to guarantee high-gain beam alignment to user $k$ can be formulated as
\begin{equation} \label{user_k_comm}
	\begin{split}
		\min_{\bw_k \in {\cal C}_w,\bm\phi \in {\cal C}_{\phi}\cap{\cal C}_{\theta},q_k^{\nu}}~\left| \bm\phi^{\rm H} \bH_k \bw_k - q_kq_k^{\nu} \right|^2,
	\end{split}
\end{equation}
where $|q_k^{\nu}|=1$ is a phase component that can be changed to ensure higher optimized DoF and performance.

Although high-quality communication of user $k$ can be achieved by solving the problem \eqref{user_k_comm}, it does not manage the interference of $k$-th user's signal to other users well.
However, for multiuser communications, effective management of inter-user interference is essential. 
Therefore, we propose an effective IM approach to enhance the achievable rate, i.e.,
\begin{equation} \label{joint_optimization}
	\begin{cases}
		\min\limits_{\bw_1 \in {\cal C}_w,\bm\phi \in {\cal C}_{\phi} \cap {\cal C}_{\theta},\bq^{\nu}}\left\| \bm\phi^{\rm H} {\rm unvec}_{N,K}(\bE_1 \bw_1) - \bq\odot \bq^{\nu}\odot \be_1 \right\|_2^2, \\
		\min\limits_{\bw_2 \in {\cal C}_w,\bm\phi \in {\cal C}_{\phi} \cap {\cal C}_{\theta},\bq^{\nu}}\left\| \bm\phi^{\rm H} {\rm unvec}_{N,K}(\bE_1 \bw_2) - \bq\odot \bq^{\nu}\odot \be_2 \right\|_2^2, \\
		\dots\dots \\
		\min\limits_{\bw_K \in {\cal C}_w,\bm\phi \in {\cal C}_{\phi} \cap {\cal C}_{\theta},\bq^{\nu}}\left\| \bm\phi^{\rm H} {\rm unvec}_{N,K}(\bE_1 \bw_K) - \bq\odot \bq^{\nu}\odot \be_K \right\|_2^2,
	\end{cases}
\end{equation}
where $\bE_1 = \left[ \bG^{\rm T}\diag(\bh_1^{*}), \dots, \bG^{\rm T}\diag(\bh_K^{*}) \right]^{\rm T} \in \mathbb{C}^{NK\times M}$; $\bq = \left[ q_1,\dots,q_K \right]^{\rm T}$ and $\bq^{\nu} = \left[ q_1^{\nu},\dots,q_K^{\nu} \right]^{\rm T}$ denote the amplitude gain and beam phase shift of all $K$ users, respectively;
$\be_k = \left[ 0,\dots,1,\dots,0 \right]^{\rm T}$ is the indicator vector we define to manage inter-user interference, where only the $k$-th element is 1 and others are 0.

However, although we can reduce the inter-user interference well by solving the distributed problem \eqref{joint_optimization}, this may lead to yielding multiple different optimal solutions with respect to $\bm\phi$, which will result in an ineffective configuration of the RIS phase shift in practice.
Therefore, we should find a compromise to obtain effective phase shift and precoding configurations to manage inter-user interference.
Based on this, we reformulate the distributed problem \eqref{joint_optimization} as a joint optimization problem, which ensures uniform variable optimization to achieve IM, given by
\begin{equation}\label{multiuser_comm_design}
	\begin{split}
		\min_{ \bW, \bm\phi, \bQ^{\nu} }&~ \left\| \bm\phi^{\rm H}\, \bXi - {\rm vec}^{\rm T}(\bQ \odot \bQ^\nu) \right\|_2^2 \\[-3pt]
		\mbox{s.t.}&~ \bW \in {\cal C}_w,\, \bm\phi \in {\cal C}_{\phi}\cap{\cal C}_{\theta},\, {\cal C}_q: |Q_{i,m}^{\nu}| = 1,
	\end{split}
\end{equation}
for $i,m=1,\dots,K,$ where $\bXi = {\rm unvec}_{N,K^2} \left({\rm vec}(\bE_1 \bW)\right)$,
$\bW = \left[ \bw_1,\dots,\bw_K \right]$, $\bE_1\bW \in \mathbb{C}^{NK\times K}$, and $\bQ\in \mathbb{C}^{K\times K}$ denotes the desired gain matrix.

It should be emphasized that the implication of the above problem is to achieve an approximation of the desired gain matrix by optimizing $\bW$ and $\bm\phi$.
To explain it more intuitively, we take the example of a 2-user system, i.e., $\bW = \left[ \bw_1,\bw_2 \right]$.
For effective inter-user IM, we expect to implement the following
\begin{equation}
	\left[
	\begin{array}{cc}
		\bm\phi^{\rm H} \bH_1 \bw_1 & \bm\phi^{\rm H} \bH_2 \bw_1 \\
		\bm\phi^{\rm H} \bH_1 \bw_2  & \bm\phi^{\rm H} \bH_2 \bw_2
	\end{array}
	\right] \rightarrow 
	\left[
	\begin{array}{ccc}
		Q_{1,1} & Q_{2,1} \\
		Q_{1,2}  & Q_{2,2} 
	\end{array}
	\right],
\end{equation}
where $|Q_{1,1}| > 0$, $|Q_{2,2}| > 0$, $|Q_{1,2}| \rightarrow 0$, and $|Q_{2,2}| \rightarrow 0$.

The above problem can also be solved by the AO algorithm.
The subproblem with respect to $\bm\phi$ can be solved by the IPDD algorithm.
However, the subproblem with respect to $\bW$ is very tricky due to the implicit variable $\bW$.
Therefore, we convert it into an optimization problem with an explicit representation of variable $\bW$ as follows
\begin{equation}\label{multiuser_w}
	\begin{split}
		\min_{ \bw } \left\| ( \bI_K \otimes \bF ) \bw - \widetilde\bq \right\|_2^2 ~~~
		\mbox{s.t.}	~ \left\|\bw \right\|_2^2 \leq P_{\rm max}, 		
	\end{split}
\end{equation}
where $\widetilde\bq = {\rm vec}^{\rm T}(\bQ \odot \bQ^\nu)$, $\bF = \left({\rm unvec}_{M,K}\left( \bE_2^{\rm T} \bm\phi^{*}  \right) \right)^{\rm T}$, and $\bE_2 = \left[ \bH_1, \dots, \bH_K \right]$.

By using the Lagrangian method, the optimal closed-form solution of the problem \eqref{multiuser_w} can be derived as \vspace{-0.2cm}
\begin{equation}
	\begin{split}
		\bW^{\star} = {\rm unvec}_{M,K}\left( \left( \widetilde\bF^{\rm H} \widetilde\bF + \gamma \bI_M \right)^{-1} \widetilde\bF^{\rm H} \widetilde\bq \right),
	\end{split}
\end{equation}
where $\widetilde\bF = (\bI_K \otimes \bF)$ and $\gamma$ can be found by the one-dimensional line search.

\vspace{-0.2cm}
\subsection{Desired Gain Matrix Adaptive Adjustment }
To achieve better inter-user fairness, we introduce Jain's fairness index in the iteration for fairness evaluation, which is given by \vspace{-0.2cm}
\begin{equation}
	\begin{split}
		J = \frac{ (\sum_{k=1}^{K} R_k )^2 }{K\cdot \sum_{k=1}^{K} R_k^2 },
	\end{split}
\end{equation}
where $J\in [1/K,1]$. $J=1$ indicates complete fairness (all users get the same transmission rate), and $J=1/K$ indicates complete unfairness (only one user gets all resources).
We expect to design the desired gain matrix to adaptively adjust to enhance Jain's fairness index $J$ in iterations.

Then, we first design each parameter of the desired gain matrix based on the channel qualities.
For the diagonal elements of $\bQ$ (desired signal gain control), we introduce a channel quality compensation mechanism:
\begin{equation}
	\begin{split}
		Q_{k,k} = \alpha \cdot \left( \frac{\min_k \| \bH_k \|_{\rm F}}{\|\bH_k \|_{\rm F}} \right)^{\gamma_1},
	\end{split}
\end{equation}
where 
$\alpha \geq 0$ is the parameter that controls the overall desired signal amplitude, and $\gamma_1$ is the compensation factor, which controls the degree of compensation.

For the non-diagonal elements of $\bQ$ (interference signal gain control), we introduce dynamic adjustment based on channel conditions:
\begin{equation}
	\begin{split}
		Q_{i,k} = \beta \cdot \left( \frac{\|\bH_k \|_{\rm F}}{\|\bH_i \|_{\rm F}+ \epsilon_h}  \right)^{\gamma_2} \cdot (1+\varrho_{i,k})^{-\gamma_3}, i\neq k,
	\end{split}
\end{equation}
where $\beta \geq 0$ is the parameter that controls the overall intensity of the interference; 
$\epsilon_h$ is a small positive number that prevents the denominator from being zero;
$\gamma_2$ and $\gamma_3$ are the interference adjustment factor;
$\varrho_{i,k}$ is the channel correlation factor and 
$$\varrho_{i,k} = \frac{ \left| {\rm vec}^{\rm H}(\bH_i) {\rm vec}(\bH_k) \right| }{\left\| \bH_i \right\|_{\rm F} \left\| \bH_k \right\|_{\rm F}  }.$$

\vspace{-0.2cm}
During adaptive adjustment, we can realize the enhancement of $J$ by changing the parameters $\chi \triangleq \left\{ \alpha, \beta, \gamma_1, \gamma_2, \gamma_3 \right\}$.
However, these parameters are changed by changing the optimal solution of the problem~\eqref{multiuser_comm_design} to $J$, which is an implicit change relation. 
Therefore, we use a forward difference numerical gradient to find the gradient of $J$ with respect to the parameters $\chi$, thus realizing the enhancement of $J$ through the gradient ascent of the parameters.
The numerical gradient is given by
\begin{equation}
	\begin{split}
		\nabla_{\chi} J(\chi) = \frac{J(\chi+\Delta \chi) - J(\chi) }{\Delta \chi},
	\end{split}
\end{equation}
where $\Delta \chi$ denotes the small perturbation applied to the parameter.
Note that the above numerical gradient is solved for each parameter in the parameter set $\chi$.

Then, we can use the gradient ascent algorithm to update the parameter
\[
\chi = \chi + \eta_\chi \nabla_\chi J(\chi),
\]
where $\eta_\chi$ denotes the ascent step length.

\vspace{-0.25cm}
\subsection{Closed-Form Method for RIS Phase-Shift Optimization }
In order to nearly reduce the complexity of RIS phase-shift optimization, we propose a completely closed-form method (CFM) that can directly obtain the sub-optimal solution through a single computation of the closed-form solution.
Note that this method needs to first compute the correlation between different user channels. Specifically, we can rewrite the objective function of the phase-shift optimization problem as \vspace{-0.2cm}
\begin{equation*}
	\begin{split}
		&\left\| \bm\phi^{\rm H} \bXi - \widetilde\bq^{\rm T}  \right\|_2^2 = \sum_{i=1}^{K^2} \left| \bm\phi^{\rm H}\bXi_{:,i} - \widetilde q_i \right|^2 \\ 
		&= \sum_{i=1}^{K^2} \left\{ \left| \sum_{n=1}^{N} \phi_n^* \Xi_{n,i} \right|^2 + |\widetilde q_i|^2 - 2 {\rm Re}\left\{ \sum_{n=1}^{N} \phi_n^* \Xi_{n,i} \widetilde q_i^* \right\}  \right\},
	\end{split}
\end{equation*}
where \vspace{-0.2cm}
\begin{equation*}
	\begin{split}
		\left| \sum_{n=1}^{N} \phi_n^* \Xi_{n,i} \right|^2 = \sum_{n=1}^{N} | \Xi_{n,i} |^2 + \sum_{n=1}^{N} \sum_{m\neq n}^{N} \phi_n^* \phi_m \Xi_{n,i} \Xi_{m,i}^*.
	\end{split}
\end{equation*}

\vspace{-0.2cm}
Based on the reformulated objective function expression, the optimal solution of $\bm\phi$ depends on $\sum_{n=1}^{N} \phi_n^* \Xi_{n,i} \widetilde q_i^*$ and the cross term $\sum_{n=1}^{N} \sum_{m\neq n}^{N} \phi_n^* \phi_m \Xi_{n,i} \Xi_{m,i}^*$. 
It is important to note that the cross term is obtained based on the channel and the precoding, then when the correlation between the channels is smaller, the effect of the cross term on $\bm\phi$ will be smaller. 
Existing studies~\cite{Dong2022NFChannel} have shown that the correlation between multiuser NF channels decreases sharply with increasing user distance. 
Moreover, in practice, we can first evaluate the channel correlation and if the correlation is within tolerable limits, then the effect of the cross terms can be ignored.
Then, we can obtain the solution of the phase shift $\phi_n$ is given by
\begin{equation} 
	\phi_n^{\star}  = 
	\begin{cases}
		e^{j\frac{(\xi_n-1)\pi}{2^{v-1}}}, ~\mbox{if}~ \angle(\sum_{i=1}^{K^2} \Xi_{n,i} \widetilde q_i^*) < \frac{(2 \xi_n - 1)\pi}{2^{v}}, \\
		e^{j\frac{\xi_n\pi}{2^{v-1}}} , ~~~\mbox{if}~ \angle(\sum_{i=1}^{K^2} \Xi_{n,i} \widetilde q_i^*) \geq \frac{(2 \xi_n - 1)\pi}{2^{v}},
	\end{cases}
\end{equation}
where $\xi_n = \lfloor \frac{2^{v-1}\angle(\sum_{i=1}^{K^2} \Xi_{n,i} \widetilde q_i^*) + \pi }{\pi} \rfloor$ for $n=1,\dots,N$.

\section{Extension to Hybrid Precoding Design}
Inspired by our proposed codebook construct methods, we can formulate an effective design method to achieve a hybrid precoding codebook construct.
The transmitted signal from the BS realized by hybrid precoding is given by
\[
\bx = \bV_{\rm A} \bv_{\rm D} s,
\]
where $\bV_{\rm A} \in \mathbb{C}^{M\times M_{\rm RF}}$ and $\bv_{\rm D} \in \mathbb{C}^{M_{\rm RF} \times 1}$ denote the analog beamformer and the digital beamformer, respectively.

Then, we can design full-digital precoding vectors based on Section III and then approach the performance of full-digital precoding by hybrid precoding.
Therefore, we can formulate the multi-resolution codebook construction problem of hybrid precoding as
\begin{equation}\label{hybrid_codebook}
	\begin{split}
		&\min_{ \bV_{\rm A},\, \bv_{\rm D} }~ \left\| \bV_{\rm A} \bv_{\rm D}  - \bw_t^{\star} \right\|_2^2 \\
		\mbox{s.t.}	
		&~ \left\| \bV_{\rm A} \bv_{\rm D} \right\|_2^2 \leq P_{\rm max},~ [\bV_{\rm A}]_{n,i} = e^{j \vartheta_{n,i}},\\
		&~ \vartheta_{n,i} \in \left\{ 0, \frac{2\pi}{2^v},\dots,\frac{(2^v-1)2\pi}{2^v} \right\},\, n = 1,\dots,N,
	\end{split}
\end{equation}
where $\bw_t^{\star}$ is obtained by solving the problem~\eqref{codeword_design_multi_antenna}.

{\it 1) Optimize $\bv_{\rm D}$ }

The problem with respect to $\bv_{\rm D}$ can be reformulated as
\begin{equation}\label{optimize_v_D}
	\begin{split}
		\min_{  \bv_{\rm D} }~ \left\| \bV_{\rm A} \bv_{\rm D}  - \bw_t^{\star} \right\|_2^2 
		~~\mbox{s.t.}	
		~ \left\| \bV_{\rm A} \bv_{\rm D} \right\|_2^2 \leq P_{\rm max},
	\end{split}
\end{equation}
which can be transformed into
\begin{equation}\label{optimize_v_D_transform}
	\begin{split}
		\min_{  \bv_{\rm D} }&~ \bv_{\rm D}^{\rm H} \bV_{\rm A}^{\rm H} \bV_{\rm A} \bv_{\rm D} - 2{\rm Re}\left\{ (\bw_t^{\star})^{\rm H} \bV_{\rm A} \bv_{\rm D}  \right\}     \\
		\mbox{s.t.}	
		&~ \bv_{\rm D}^{\rm H}  \bV_{\rm A}^{\rm H} \bV_{\rm A} \bv_{\rm D} \leq P_{\rm max}.
	\end{split}
\end{equation}

By the Lagrangian method, the optimal solution of the problem~\eqref{optimize_v_D_transform} is given by
\begin{equation}
	\begin{split}
		\bv_{\rm D}^{\star} &= (\bV_{\rm A}^{\rm H} \bV_{\rm A} )^{-1}\bV_{\rm A}^{\rm H}\bw_t^{\star}  \big/ (1 + \lambda_v), \\
		\lambda_v &= \sqrt{ \frac{(\bw_t^{\star})^{\rm H}\bV_{\rm A} (\bV_{\rm A}^{\rm H} \bV_{\rm A} )^{-1} \bV_{\rm A}^{\rm H} \bw_t^{\star} }{P_{\rm max}} } - 1.
	\end{split}
\end{equation}

\vspace{-0.2cm}
{\it 2) Optimize $\bV_{\rm A}$ }

The problem with respect to $\bV_{\rm A}$ can be reformulated as
\begin{equation*}\label{hybrid_codebook}
	\begin{split}
		\min_{ \bV_{\rm A} }&~ \bv_{\rm A}^{\rm H} \left( \bv_{\rm D}^* \bv_{\rm D}^{\rm T} \otimes \bI \right) \bv_{\rm A} - 2 {\rm Re}\left\{ {\rm vec}^{\rm H}( \bw_t^{\star} \bv_{\rm D}^{\rm H} ) \bv_{\rm A} \right\} \\
		\mbox{s.t.}	
		&~ \bv_{\rm A}^{\rm H} \left( \bv_{\rm D}^* \bv_{\rm D}^{\rm T} \otimes \bI \right) \bv_{\rm A} \leq P_{\rm max}, \\
		&~ [\bV_{\rm A}]_{n,i} = e^{j \vartheta_{n,i}},\\
		&~ \vartheta_{n,i} \in \left\{ 0, \frac{2\pi}{2^v},\dots,\frac{(2^v-1)2\pi}{2^v} \right\},\, n = 1,\dots,N,
	\end{split}
\end{equation*}
where $\bv_{\rm A} = {\rm vec}(\bV_{\rm A})$. This problem can likewise be solved using the same algorithm as in Section IV-B.

\vspace{-0.35cm}
\section{Simulation Results}
In this section, we demonstrate the effectiveness of the proposed multi-resolution codebook design and multiuser IM method by computer simulations.
We set the noise power $\sigma_k^2$ as -110dBm, the carrier frequency as 10GHz, the coordinate of the BS as $(-40,0,-25)$ meters, the coordinate of the RIS as $(0,0,0)$ meters, the number of RIS elements as $N_1 =128$ and $N_2 = 4$, and the RIS array aperture as $D_{\rm ris} = \sqrt{(127\lambda_c/2)^2 + (3\lambda_c/2)^2} \approx 1.9 \,$m \cite{Wei2022Chinacomm}.
All users are located in the $x$-$z$ plane with $y_u=0$ and the sampling range is set from $-1000\lambda_c$ to $1000\lambda_c$ in the $x$-axis and from $500\lambda_c$ to $2500\lambda_c$ in the $z$-axis.
According to \cite{Cui2023CommunMag}, the Rayleigh distance $\frac{2D_{\rm ris}^2}{\lambda_c}$ is larger than $\frac{r_{\rm br}r_{\rm ru}}{r_{\rm br} + r_{\rm ru}} $ causing both the BS and the user to be in the NF range, where $r_{\rm br}$ and $r_{\rm ru}$ denote the distance from the BS to the RIS and from the RIS to the user, respectively.
The desired beam gain $C_g$ can be set on demand, in this paper we set it to 30dB to widen the difference in beam gain between different positions.

\vspace{-0.35cm}
\subsection{Hierarchical Beam Training}
\begin{figure}[t]	
	\centering \includegraphics[width=0.7\linewidth]{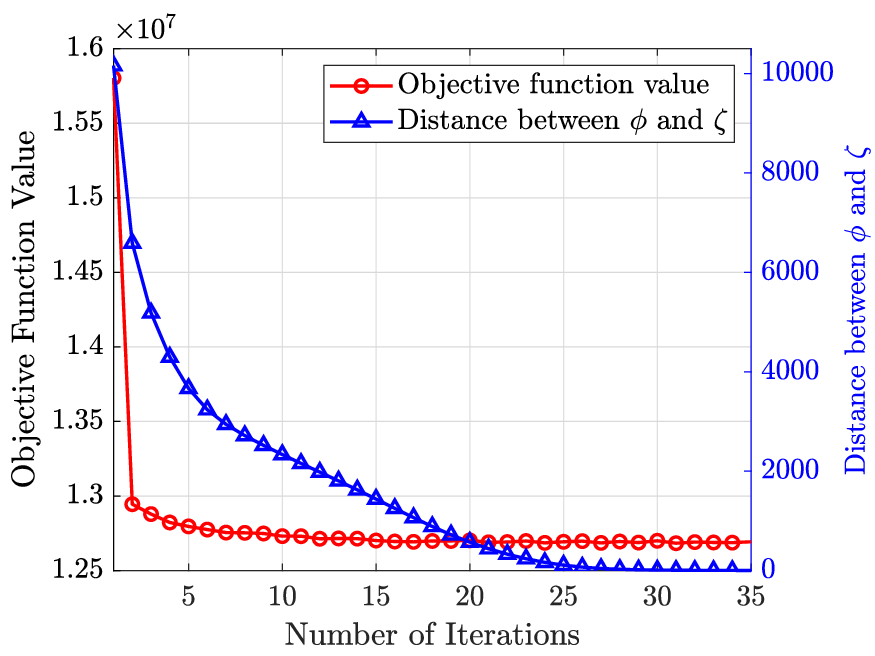}
	\vspace{-0.3cm}
	\caption{Convergence of our proposed algorithm, $M=4, v=2$, SNR = 6dB.}
	\vspace{-0.7cm}
	\label{fig:Algorithm_convergence}
\end{figure}

\begin{figure}[t]
	\centering
	\subfloat[Desired beam pattern ]{\includegraphics[width=0.33\columnwidth]{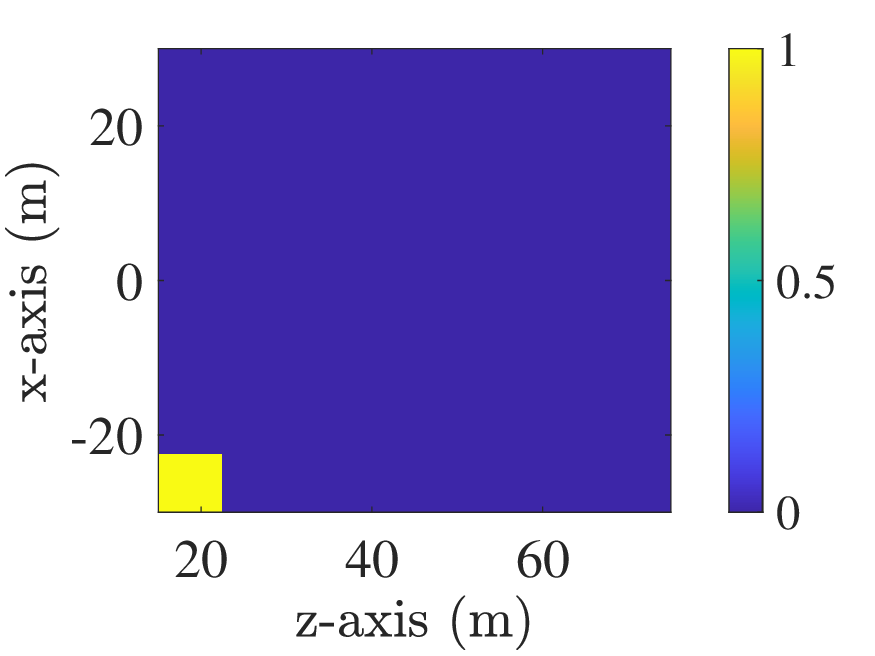}%
		\label{Plane_ideal_beam_layer1}}
	\hfil
	\subfloat[JOCC method ]{\includegraphics[width=0.33\columnwidth]{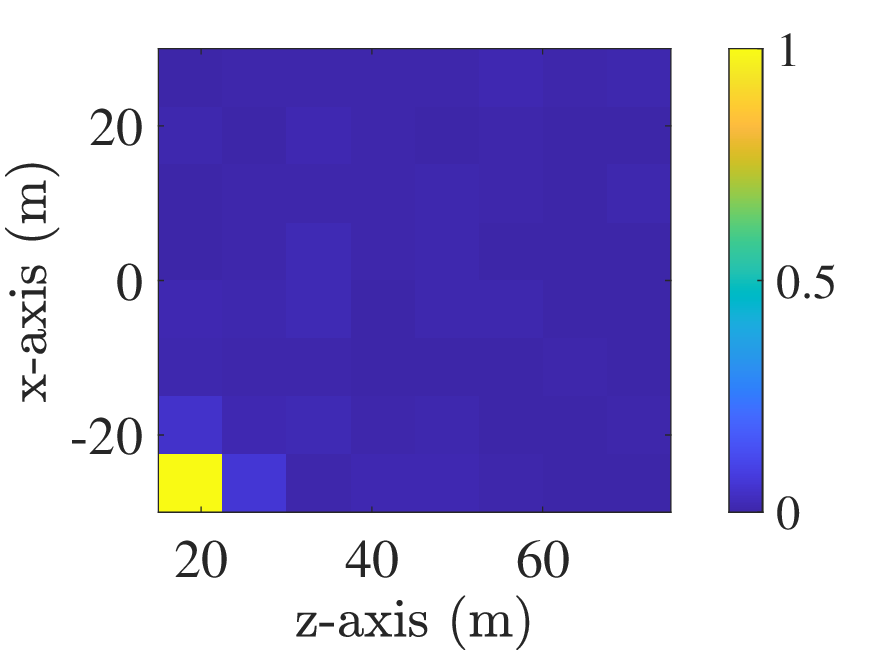}%
		\label{Plane_beam_practicalRIS_layer1}}
	\hfil
	\subfloat[SOCC method ]{\includegraphics[width=0.33\columnwidth]{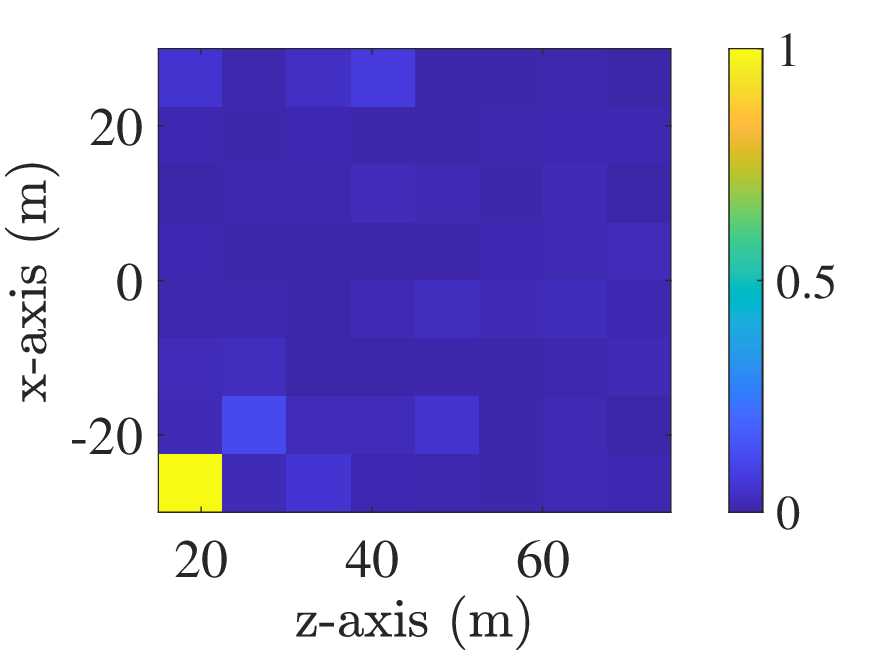}%
		\label{Plane_beam_ideal_RIS_layer1}}
	\hfil
	\vspace{-0.3cm}
	\centering
	\subfloat[Desired beam pattern ]{\includegraphics[width=0.32\columnwidth]{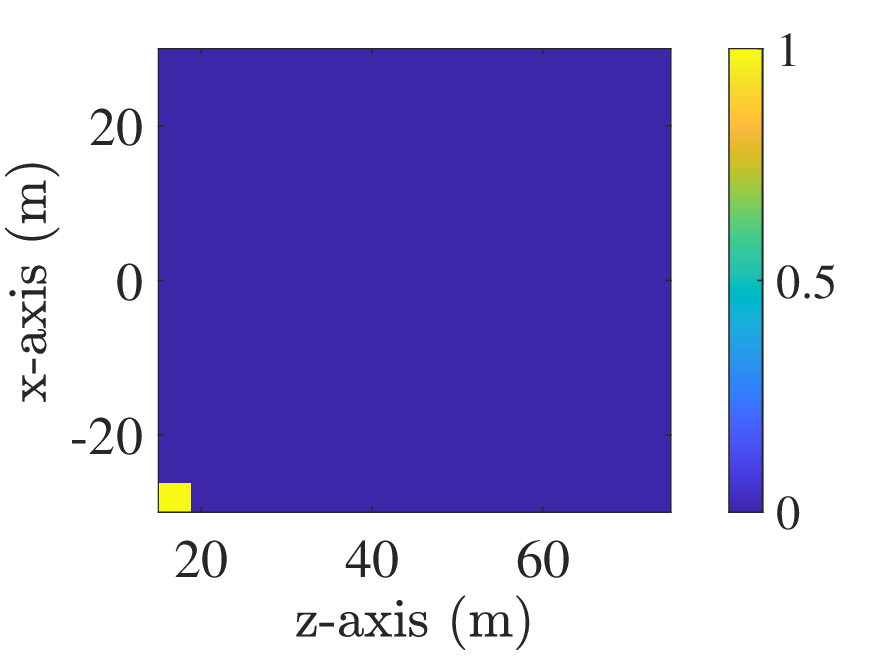}%
		\label{Plane_ideal_beam_layer2}}
	\hfil
	\subfloat[JOCC method]{\includegraphics[width=0.32\columnwidth]{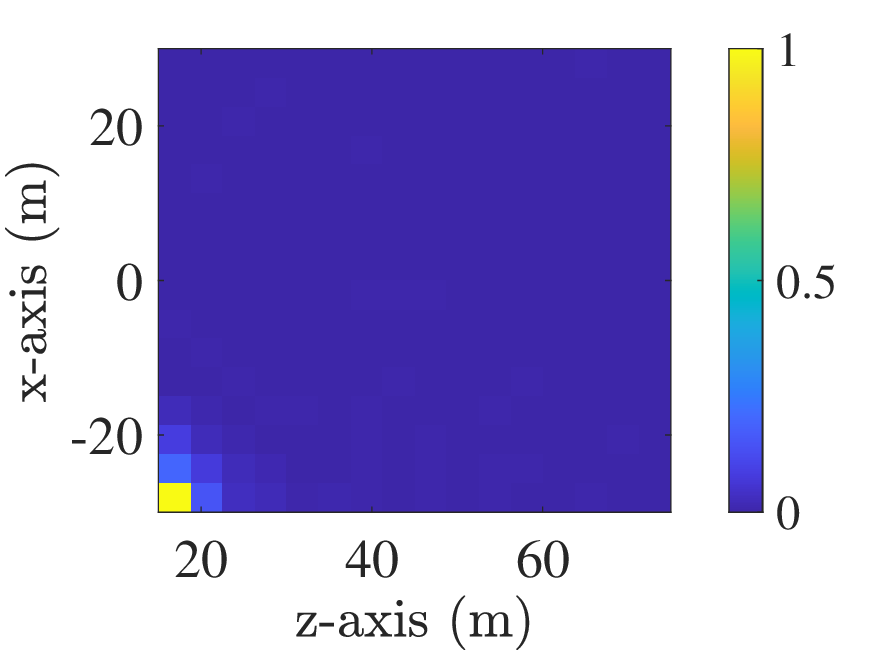}%
		\label{Plane_beam_practicalRIS_layer21}}
	\hfil
	\subfloat[SOCC method]{\includegraphics[width=0.32\columnwidth]{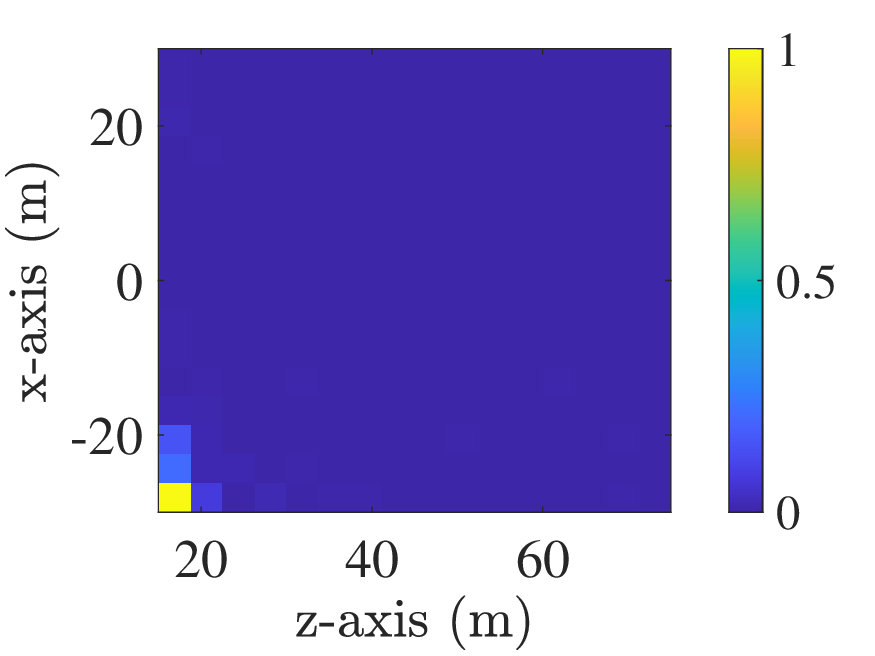}%
		\label{Plane_beam_ideal_RIS_layer2}}
	\caption{The beam pattern obtained by the proposed method under different levels with $v=2$. Level 1: (a)(b)(c), level 2: (d)(e)(f).}
	\vspace{-0.5cm}
	\label{fig:beam_pattern}
\end{figure}

In the hierarchical beam training simulation, we consider a two-level beam training and corresponding codebook design.
During the multi-resolution codebook design, the number of all sampling points along the $x$-axis and $z$-axis is set to $2^8$ and $2^5$, respectively, i.e., $S_x = 2^8$ and $S_z = 2^5$.
For the two-level beam training, the corresponding number of sampling points along the x-axis and z-axis in different levels is set to $S_1^x = 2^3$, $S_1^z = 2^2$, $S_2^x = 2^6$, and $S_2^z = 2^4$.
With this sampling setup, we first verify the convergence of the proposed optimization algorithm for multi-resolution codebook design and the approximation behavior of the continuous phase shifts to discrete phase points.
From Fig.~\ref{fig:Algorithm_convergence}, the proposed optimization algorithm ensures excellent convergence and fast convergence for the codebook design problem. 
In addition, it can also be seen that as the iteration proceeds the continuous phase shift is gradually approximated to the discrete phase shift until the two are the same, which ensures the optimization performance under the discrete phase shift of the RIS.
With the proposed algorithm, we can obtain an exact match of the desired beam pattern by either the JOCC method or the SOCC method, as shown in Fig.~\ref{fig:beam_pattern}.
\begin{figure}[t]	
	\centering \includegraphics[width=0.7\linewidth]{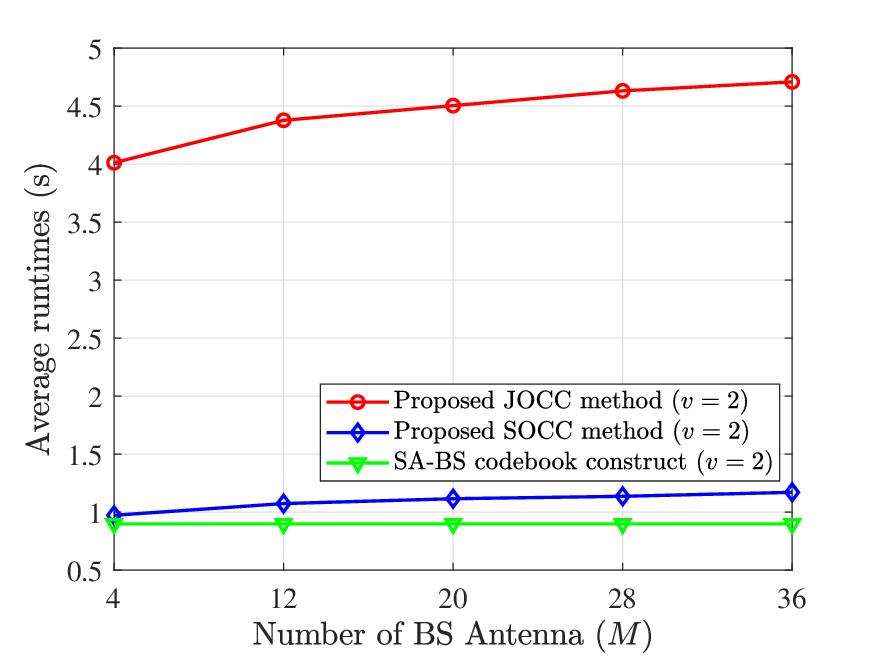}
	\vspace{-0.3cm}
	\caption{Average runtimes of the proposed different codebook construct methods, $M=4, v=2$, SNR = 6dB.}
	\vspace{-0.5cm}
	\label{fig:Codebook_construct_runtimes}
\end{figure}
\begin{figure}[t]	
	\centering \includegraphics[width=0.7\linewidth]{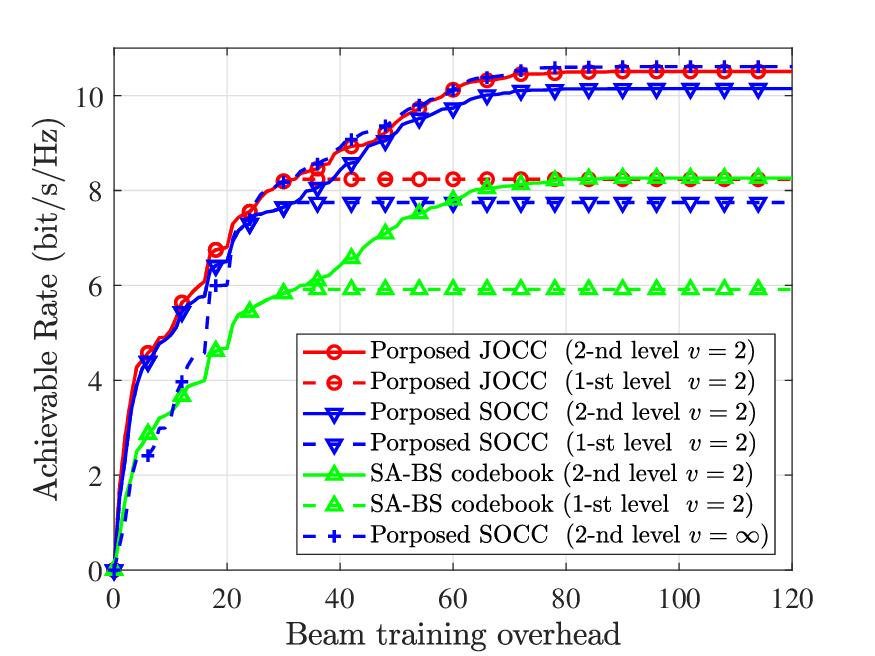}
	\vspace{-0.25cm}
	\caption{Two-level beam training performance under JOCC, SOCC and SA-BS codebook construction methods, $M=4, v=2$, SNR = 6dB.}
	\vspace{-0.5cm}
	\label{fig:Rate_beam_training_2level}
\end{figure}
\begin{figure}[h!]	
	\centering \includegraphics[width=0.7\linewidth]{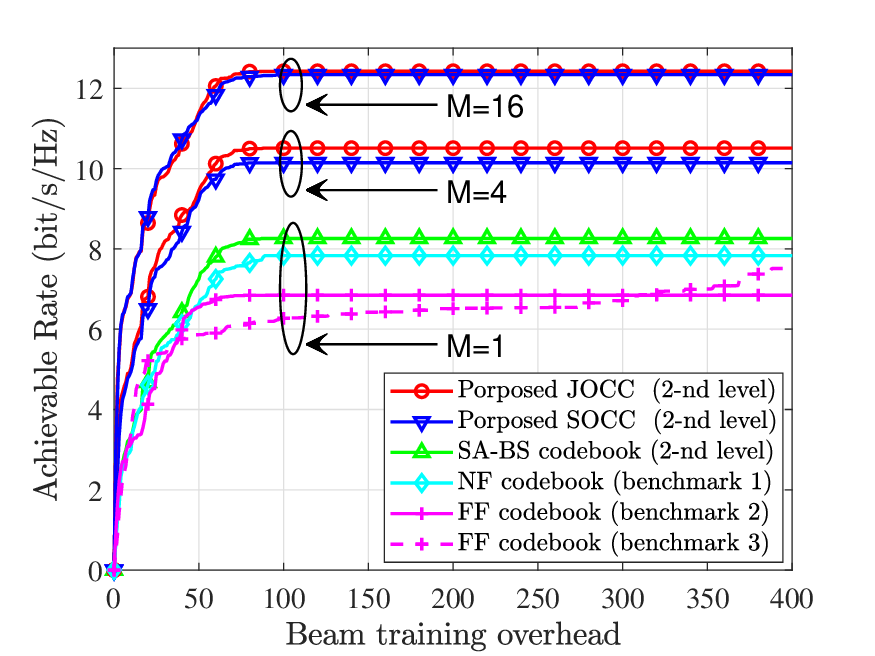}
	\vspace{-0.3cm}
	\caption{Comparison of different codebook construction methods with different number of BS antennas, $v=2$, SNR = 6dB.}
	\vspace{-0.5cm}
	\label{fig:Rate_beam_training_comparison}
\end{figure}

Further, to validate the complexity of the proposed multi-resolution codebook construction methods, we simulate the average runtime of algorithms for different numbers of BS antennas. 
As shown in Fig.~\ref{fig:Codebook_construct_runtimes}, the proposed SOCC method can greatly reduce the design complexity compared to the proposed JOCC method, and the complexity of the SOCC method is close to that of the codebook design method in SA-BS systems.
Although the complexity is close, the performance of the proposed SOCC method is vastly improved compared to the SA-BS codebook in beam training, as shown in Fig.~\ref{fig:Rate_beam_training_2level}.
We can also see that further performance improvement can be achieved by performing the second level of beam training within the range obtained from the first level of beam training, which further demonstrates that the proposed multi-resolution codebook construction methods can accurately achieve the matching of user positions. This is intertwined with Fig.~\ref{fig:beam_pattern}.

Moreover, in order to verify the superiority of the proposed codebook construction methods in hierarchical beam training, we compare the proposed method with codebook design methods based on existing research ideas, and the benchmark methods are as follows.

\noindent
\noindent 1) benchmark 1: the NF codebook designed based on the idea of existing research~\cite{Wei2022Chinacomm,Lv2024codebook} is applied in an XL-RIS-aided SA-BS system with a 2-bit phase shift;

\noindent
\noindent 2) benchmark 2: the FF multi-resolution codebook designed based on the idea of existing research~\cite{Noh2017FFHL} is applied in an XL-RIS-aided SA-BS NF system with a 2-bit phase shift;

\noindent
\noindent 3) benchmark 3: the FF exhaustive search codebook designed based on the idea of existing research~\cite{Lee2016FFES} is applied in an XL-RIS-aided SA-BS NF system with a 2-bit phase shift.

\noindent
From Fig.~\ref{fig:Rate_beam_training_comparison}, it can be seen that the proposed method can obtain a higher achievable rate even in the SA-BS system than that obtained by benchmark methods.
In addition, benchmark 1 for matching the NF beam training cannot provide guidance for the design of MA-BS systems.
Compared with the benchmark methods, our proposed JOCC method and SOCC method for MA-BS systems achieve significant performance improvement, which is also instructive for codebook design in practical systems.

\vspace{-0.3cm}
\subsection{Multiuser Interference Management}
\begin{figure}[t]	
	\centering \includegraphics[width=0.7\linewidth]{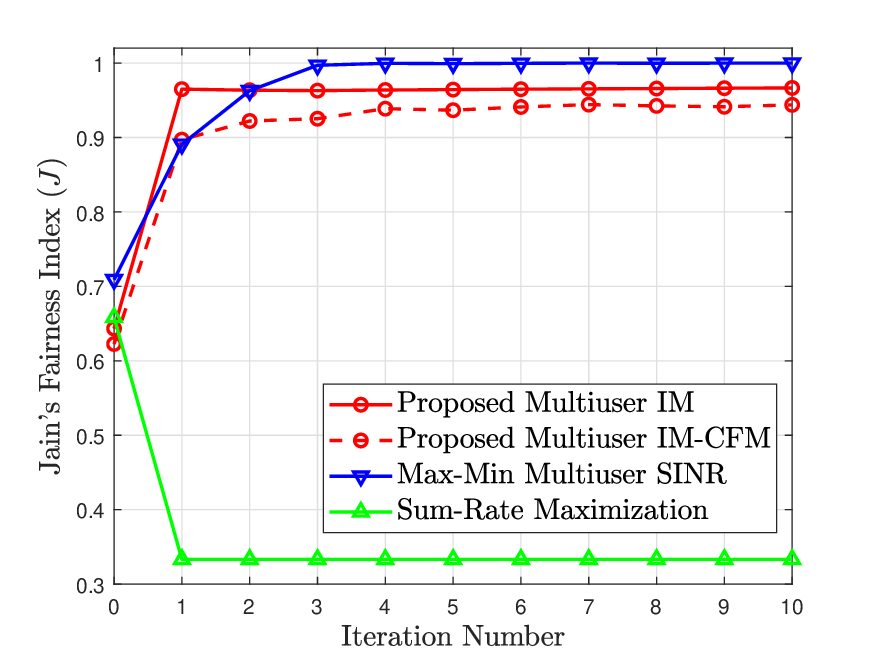}
	\vspace{-0.3cm}
	\caption{Variation of Jain's fairness index with number of iterations for different method, $M=16,N=512,K=3,v=3,P_{\rm max}=40\,$dBm. }
	\vspace{-0.5cm}
	\label{fig:Jain_fairness_index}
\end{figure}
\begin{figure}[t]	
	\centering \includegraphics[width=0.7\linewidth]{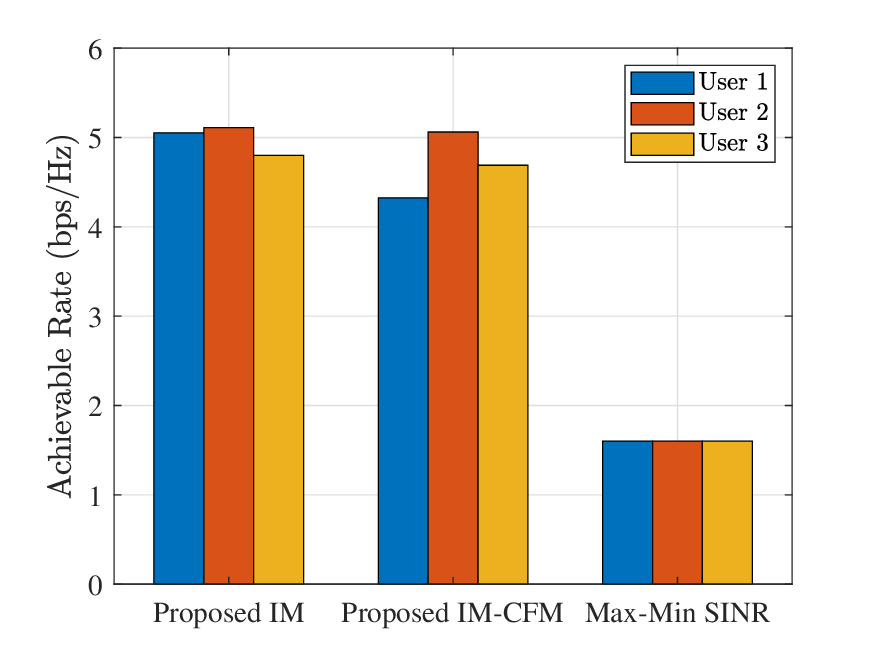}
	\vspace{-0.3cm}
	\caption{Comparison of user rates between the proposed IM and the max-min SINR, $M=16,N=512,K=3,v=3,P_{\rm max}=40\,$dBm. }
	\vspace{-0.5cm}
	\label{fig:Three_user_rate}
\end{figure}
\begin{figure}[t]	
	\centering \includegraphics[width=0.7\linewidth]{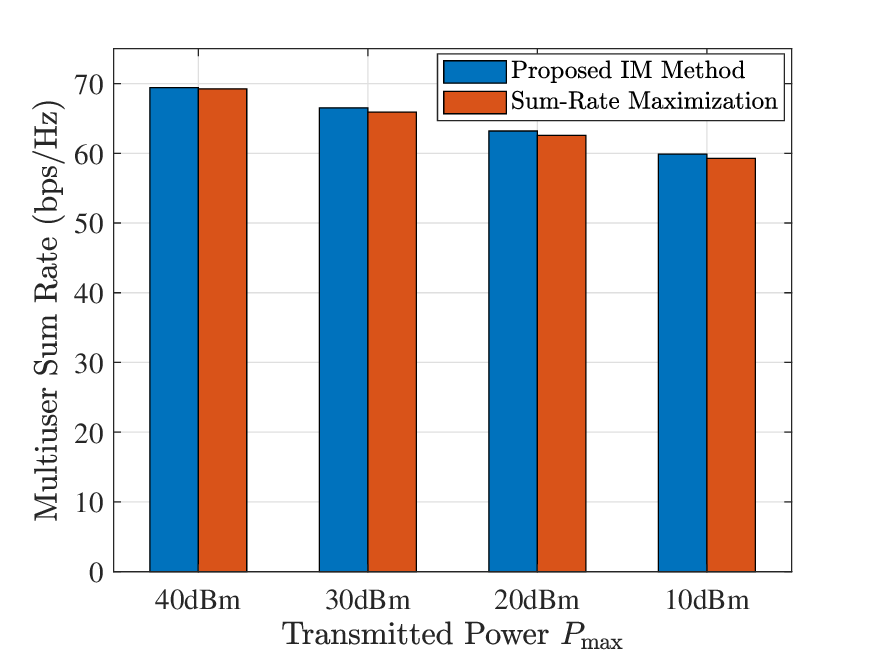}
	\vspace{-0.3cm}
	\caption{Performance comparison between the proposed IM and the sum-rate maximization at different transmit power, $M=16,N=512,K=3,v=3$. }
	\vspace{-0.5cm}
	\label{fig:Sum_rate_performance}
\end{figure}
\begin{figure}[t]	
	\centering \includegraphics[width=0.7\linewidth]{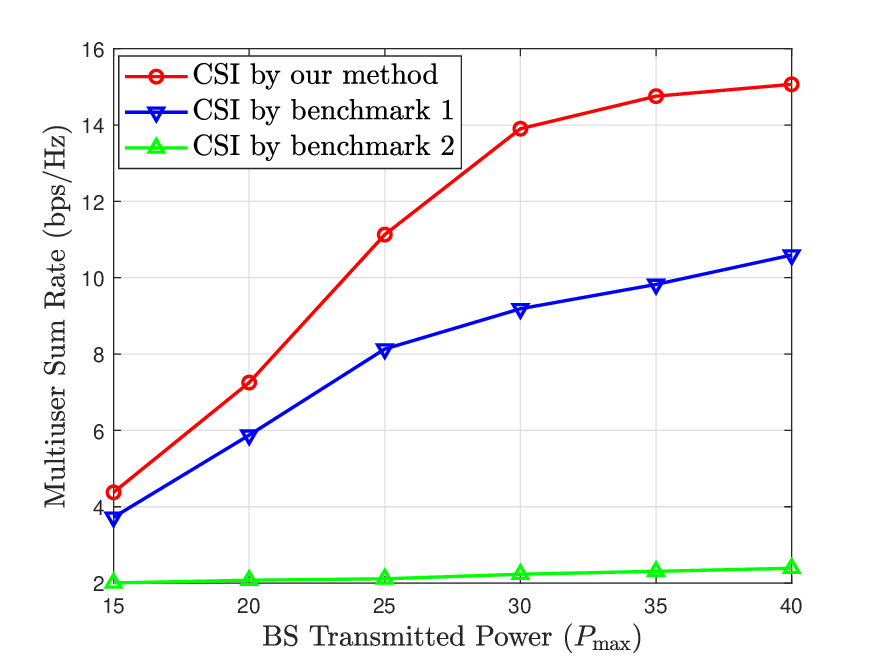}
	\vspace{-0.3cm}
	\caption{Comparison of the optimized multiuser sum rate performance based on CSI obtained by different methods, $M=16,N=512,K=3,v=3,P_{\rm max}=40\,$dBm. }
	\vspace{-0.5cm}
	\label{fig:Sum_rate_CSI}
\end{figure}
After obtaining user CSI through beam training, we can realize multiuser interference management based on CSI. 
Before that, we demonstrate the advantages of the proposed IM method over the sum-rate maximization method and the max-min multiuser SINR method through simulation.
We initialize $\alpha$, $\beta$, $\gamma_1$, $\gamma_2$, and $\gamma_3$ as $\frac{\sqrt{P_{\rm max}}}{K}$, $0.01$, 0.5, 1, and 1, respectively.
As mentioned in Remark 3, the sum-rate maximization method seeks to ensure a high rate thereby losing user fairness, and the max-min SINR method seeks extreme fairness, which can also be verified in Fig.~\ref{fig:Jain_fairness_index}.
Moreover, it can be seen that our proposed method guarantees a high level of user fairness.
The proposed CFM for optimizing $\bm\phi$ reduces the complexity while still guaranteeing a small performance loss.

From Fig.~\ref{fig:Three_user_rate}, we can see that the proposed IM method can obtain a higher achievable rate than the max-min SINR method and can maintain high-quality communication for all users. 
There are two reasons for the large performance difference between the proposed IM method and the max-min SINR method: 1) the proposed IM can trade-off the overall performance while losing a small amount of fairness while the max-min SINR method allocates more resources to users with poor channel quality; 
2) the proposed IM method has convex objective functions for $\bw$ or $\bm\phi$ and does not require the use of relaxation transformations, whereas the max-min SINR method already incurs a performance loss during relaxation.
In addition, the proposed IM method is very flexible and can achieve comparable performance to the sum rate maximization method by adjusting the desired gain matrix aligning the users with the strongest channel quality, as shown in Fig.~\ref{fig:Sum_rate_performance}.
\begin{figure}[t]	
	\centering \includegraphics[width=0.7\linewidth]{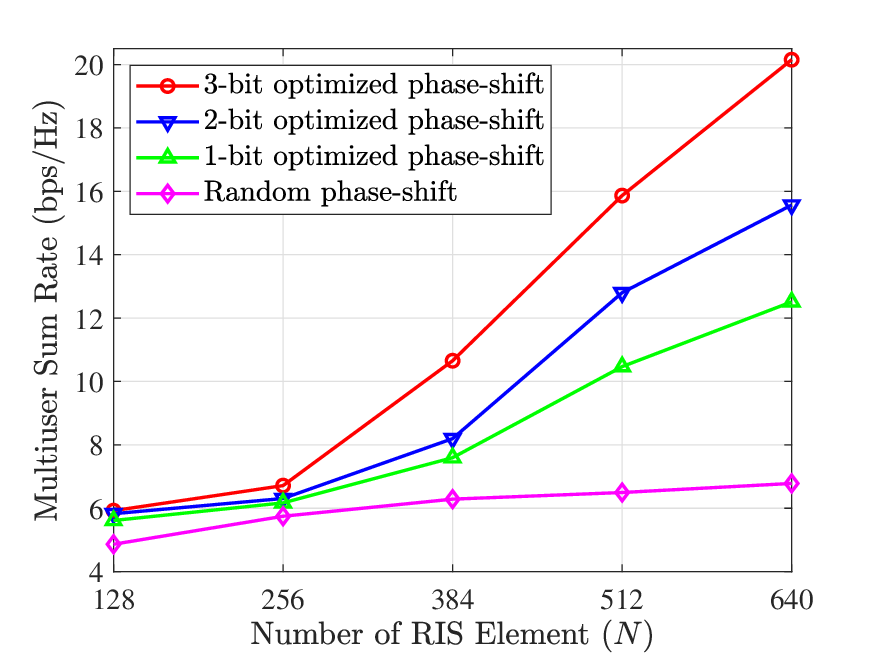}
	\vspace{-0.3cm}
	\caption{Multiuser sum rate performance under different RIS elements, $M=16,K=3,v=3,P_{\rm max}=40\,$dBm. }
	\vspace{-0.5cm}
	\label{fig:Sum_rate_N}
\end{figure}
\begin{figure}[t]	
	\centering \includegraphics[width=0.7\linewidth]{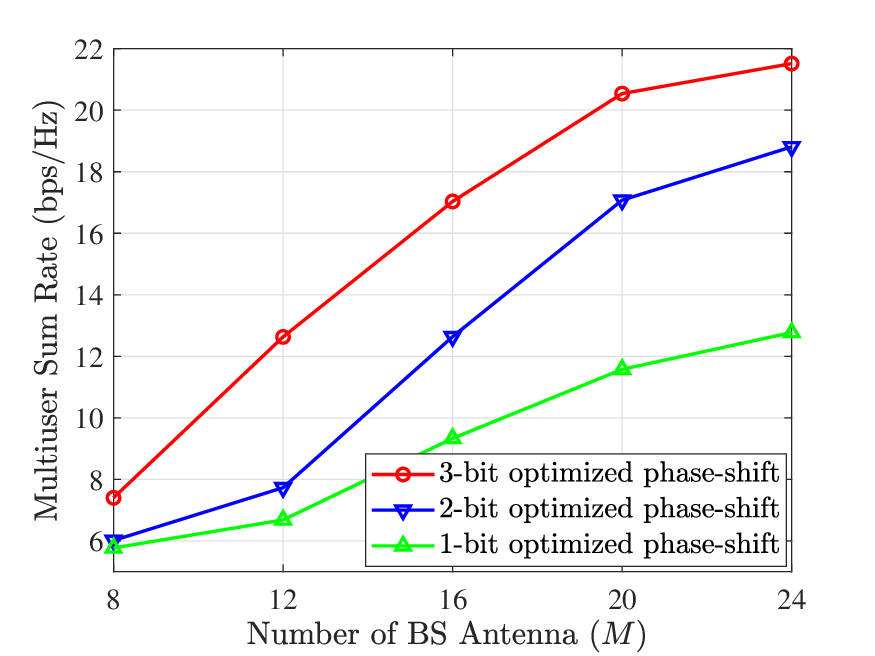}
	\vspace{-0.3cm}
	\caption{Multiuser sum rate performance under different BS antennas, $N=512,K=3,v=3,P_{\rm max}=40\,$dBm. }
	\vspace{-0.5cm}
	\label{fig:Sum_rate_M}
\end{figure}
\begin{figure}[h!]	
	\centering \includegraphics[width=0.7\linewidth]{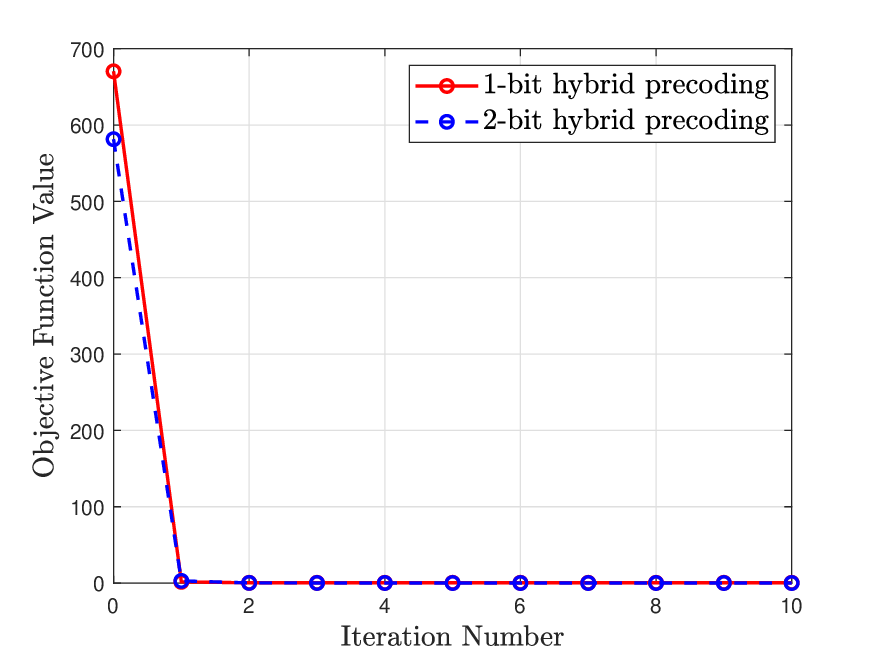}
	\vspace{-0.3cm}
	\caption{Convergence of the proposed hybrid precoding method, $M=16,M_{\rm RF} = 4,P_{\rm max}=40\,$dBm. }
	\vspace{-0.7cm}
	\label{fig:Convergence_hybrid_precoding}
\end{figure}
\begin{figure}[h!]
	\centering
	\subfloat[Desired beam pattern ]{\includegraphics[width=0.4\columnwidth]{Plane_ideal_beam_layer2.eps}%
		\label{Plane_ideal_beam_layer2_hybrid}}
	\hfil
	\subfloat[Hybrid precoding method ]{\includegraphics[width=0.4\columnwidth]{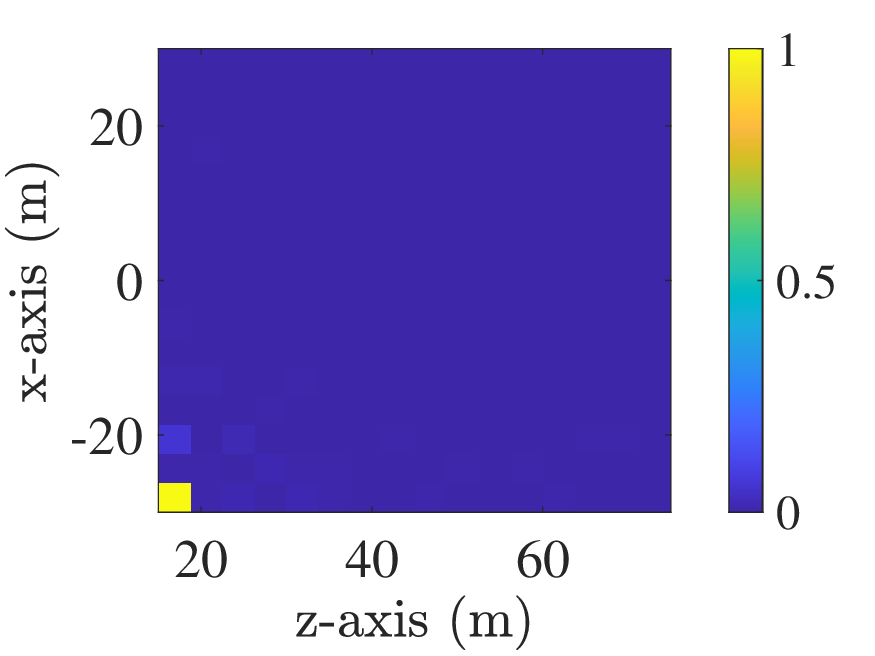}%
		\label{Plane_beam_hybrid_precoding}}
	\hfil
	\centering
	\caption{Beam pattern obtained by the proposed hybrid precoding method with $v=2$.}
	\vspace{-0.6cm}
	\label{fig:beam_pattern_hybrid}
\end{figure}

Demonstrating the superiority of the proposed IM method, we can achieve the inter-user IM based on the CSI obtained from different methods. 
As shown in Fig.~\ref{fig:Sum_rate_CSI}, the CSI obtained based on the proposed method can obtain higher sum-rate performance than the CSI obtained based on the benchmark methods, which is due to the fact that the proposed multi-resolution codebook design method can better match the hierarchical beam training and thus obtain more accurate CSI.
Further, it can be seen from Fig.~\ref{fig:Sum_rate_N} that the proposed RIS discrete phase shift optimization obtains a significant performance improvement over the random phase shift setup, and the performance further increases with the increase in the number of bits. 
Moreover, as the number of RIS reflecting elements increases, the greater performance difference at different numbers of bits, which is due to the fact that more RIS reflecting elements enable more accurate beam focusing.
Similarly, as the number of BS antennas increases, the higher beam focusing capability and the resolution between multiuser interference, and therefore the multiuser sum rate performance is further improved, as shown in Fig.~\ref{fig:Sum_rate_M}.

\vspace{-0.3cm}
\subsection{Hybrid Precoding Optimization}
In order to reduce the RF chains hardware overhead at the BS, we can use hybrid precoding of digital and analog precoding to approach the performance of full-digital precoding.
In Fig.~\ref{fig:Convergence_hybrid_precoding}, we use digital precoding with 4 RF chains and analog precoding with 2-bit phase shift to approximate the codeword of full-digital precoding with 16 RF chains, and it can be seen that the degree of approximation by both of our proposed methods is almost identical.
In this case, exact matching of the ideal beam pattern can still be achieved by hybrid precoding at the BS, as shown in Fig.~\ref{fig:beam_pattern_hybrid}.

\vspace{-0.3cm}
\section{Conclusion}
In this paper, we proposed two effective multi-resolution codebook design methods, JOCC and SOCC, tailored for the hierarchical beam training scheme in discrete XL-RIS-aided NF communication systems.
We further proposed a multiuser IM method that can be solved simply and efficiently taking into account user fairness.
In addition, we also provide effective algorithms for multiuser sum-rate maximization and max-min SINR problems.
To address the resulting non-convex optimization problems, we developed an efficient AO algorithm capable of obtaining stationary solutions to the original non-convex problem by solving each subproblem in closed-form iterations.
Simulation results demonstrate that the proposed methods significantly outperform existing methods in terms of performance. 
In addition, the extension of hybrid precoding provides ideas to reduce the hardware overhead at the BS while guaranteeing effective multi-resolution codebook construction. 
Among the multiuser IM methods, the numerical gradient ascent method as well as the RM method of optimized $\bm\phi$ provide more important guidance for practical systems.

%%%%%%%%%% Appendix %%%%%%%%%%%%%%
%\clearpage
%\ninept
%\appendix
\begin{appendices}
	\vspace{-0.3cm}
	\section*{Appendix A: Proof of Theorem 1}
	We can derive the projection $\Pi_{{\cal C}_v}(\kappa) $ by solving the problem
	\begin{equation} \label{projection_mean}
		\begin{split}
			\min_{\tau}&~ \left( {\rm Re}\{\tau\} - {\rm Re}\{\kappa\} \right)^2 + \left( {\rm Im}\{\tau\} - {\rm Im}\{\kappa\} \right)^2 \\
			{\rm s.t.}
			&~ \tau = e^{j\theta},~ \theta\in {\cal C}_v.
		\end{split}
	\end{equation}
	
\begin{figure}[t]
		\centering
		\vspace{-0.25cm}
		\subfloat[Structure 1]{\includegraphics[width=0.28\columnwidth]{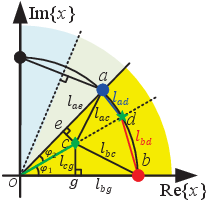}%
			\label{Discrete_ris_projection_1}}
		\hfil
		\subfloat[Structure 2 ]{\includegraphics[width=0.28\columnwidth]{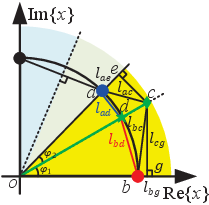}%
			\label{Discrete_ris_projection_2}}
		\hfil
		\subfloat[Structure 3 ]{\includegraphics[width=0.28\columnwidth]{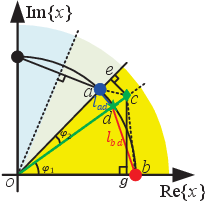}%
			\label{Discrete_ris_projection_3}}
		\vspace{-0.1cm}
		\caption{The structure of the projection from any point in the complex coordinate system to a discrete point of the unit circle.}
		\vspace{-0.4cm}
		\label{fig:Discrete_ris_projection}
	\end{figure}

	From Problem \eqref{projection_mean}, we can know that the essence of the projection $\Pi_{{\cal C}_v}(\kappa) $ is to find the closest point to the projected point in a discrete set in the two-dimensional (2D) complex plane. 
	For projection $\Pi_{{\cal C}_v}(\kappa)$, there are three structures in the 2D complex plane, as shown in~Fig.~\ref{fig:Discrete_ris_projection}.
	When $l_{ac} < l_{bc}$, the projection point of point $c$ in the discrete constraint space is point $a$.
	In these structures, the length $l_{oc}$ is known and is equal to $\sqrt{({\rm Re}\{\kappa\})^2 + ({\rm Im}\{\kappa\})^2}$.
	Therefore, the distance from point $c$ to $a$ or $b$ follows the function $y_d(\varphi)$.
	\begin{equation*}
		\begin{split}
			y_d(\varphi) &= \sqrt{ 1 + l_{oc}^2 - 2l_{oc}{\rm cos}(\varphi) },
		\end{split}
	\end{equation*}
	where $\varphi \in [0,\pi)$ must be satisfied due to the limitation from the constraint space.

%	\begin{figure}[t]	
%		\centering \includegraphics[width=0.5\linewidth]{Discrete_ris_projection_1}
%		\caption{The structure of the projection from any point in the complex coordinate system to a discrete point of the unit circle.}
%		\vspace{-0.2cm}
%		\label{fig:Discrete_ris_projection_1}
%	\end{figure}

	\begin{figure}[t]	
		\centering \includegraphics[width=0.6\linewidth]{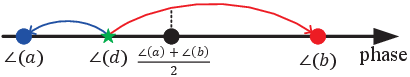}
		\vspace{-0.2cm}
		\caption{The structure of a one-dimensional phase projection.}
		\vspace{-0.5cm}
		\label{fig:1D_phase_projection}
	\end{figure}
	\begin{figure}[t]
		\centering
		\vspace{-0.25cm}
		\subfloat[Structure 1]{\includegraphics[width=0.281\columnwidth]{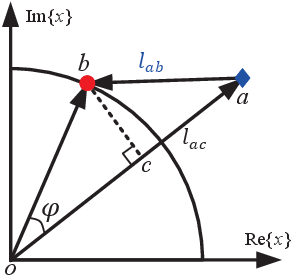}%
			\label{Phase_alignment_structure_1}}
		\hfil
		\subfloat[Structure 2]{\includegraphics[width=0.281\columnwidth]{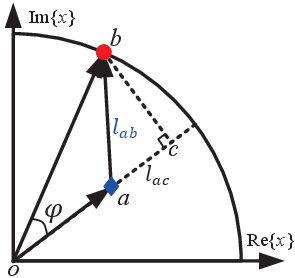}%
			\label{Phase_alignment_structure_2}}
		\vspace{-0.1cm}
		\caption{The structure of the projection from any point in the complex coordinate system to a discrete point of the unit circle.}
		\vspace{-0.53cm}
		\label{fig:Phase_alignment_structure}
	\end{figure}
	
	Further, the derivative of the function $y_d(\varphi) $ with respect to the variable $\varphi$ when $\varphi\in [0,\pi)$ is given as follows
	\begin{equation*}
		\begin{split}
			\frac{\partial y_d(\varphi)}{\partial \varphi} &= \frac{ 2 l_{oc} {\rm sin}(\varphi) }{\sqrt{ \left( 1 + l_{oc}^2 - 2 l_{oc} {\rm cos}(\varphi) \right)^2 }} \geq 0.
		\end{split}
	\end{equation*}

	Therefore, the function $y_d(\varphi)$ is a non-decreasing function with respect to $\varphi$ when $\varphi \in [0,\pi)$.
	Therefore, the projection $\Pi_{{\cal C}_v}(\kappa) $ can be further transformed into a projection operation of the point $d$ on the unit circle into the constraint space.
	Based on this, the comparison of distances in the 2D complex plane is converted into one-dimensional phase comparisons, as shown in Fig.~\ref{fig:1D_phase_projection}.
	Point $d$ is projected to point $a$ when $\angle(d)<\frac{\angle(a) + \angle(b)}{2}$, otherwise point $d$ is projected to point $b$.

%	Therefore, the projection $\Pi_{{\cal C}_v}(\kappa) $ can be equivalently transformed into the following problem
%	\begin{equation} 
%		\begin{split}
%			\min_{\theta}\, | \theta - \angle(\kappa) |^2 ~~{\rm s.t.}~ \theta\in {\cal C}_v.
%		\end{split}
%	\end{equation}
	
	Then, we assume that the phase $\angle(\kappa)$ is within the $z$-th and $(z+1)$-th points of the constraint set ${\cal C}_v$, i.e.,
	\[
	\angle(\kappa) \geq \frac{(z-1)\pi}{2^{v-1}},~
	\angle(\kappa) \leq \frac{z\pi}{2^{v-1}},
	\]
	where $z$ can be calculated as $\lfloor \frac{2^{v-1}\angle(\kappa) + \pi }{\pi} \rfloor$.

	As shown in Fig.~\ref{fig:1D_phase_projection}, we are able to obtain that the optimal $\theta$, which is equal to $\frac{(z-1)\pi}{2^{v-1}}$ when $\angle(\kappa) < \frac{(2z-1)\pi}{2^{v}}$ and $\frac{z\pi}{2^{v-1}}$ when $\angle(\kappa) \geq \frac{(2z-1)\pi}{2^{v}}$.
	
	The proof is completed. $\hfill\blacksquare$
	
	\vspace{-0.3cm}
	\section*{Appendix B: Proof of Theorem 2}
	The problem $\min\limits_{e\in \{|e|=1\}}|e - \tilde e|^2$ can be transformed into
	\begin{equation} \label{e_projection}
		\min\limits_{e\in \{|e|=1\}} \left({\rm Re}\{e - \tilde e\}\right)^2 + \left({\rm Im}\{e - \tilde e\}\right)^2.
	\end{equation}
	
	The essence of the above problem is to find a coordinate point $\left({\rm Re}\{e\}, {\rm Im}\{e\}\right)$ on the complex unit circle that is closest to the coordinate point $\left({\rm Re}\{\tilde e\}, {\rm Im}\{\tilde e\}\right)$.
	In the complex plane, the problem has two structures in the complex plane, as shown in Fig.~\ref{fig:Phase_alignment_structure}, where the coordinates of point $a$ and $b$ are $\left({\rm Re}\{\tilde e\}, {\rm Im}\{\tilde e\}\right)$ and $\left({\rm Re}\{e\}, {\rm Im}\{e\}\right)$, respectively.
	The following relationship is satisfied in both structures.
	\[
	l_{ab}(\varphi) = \sqrt{|\tilde e|^2 + 1 - 2|\tilde e|{\rm cos}(\varphi) },
	\]
	where $\varphi = |\angle(a) - \angle(b)|$, where $\angle(a)$ and $\angle(b)$ denote the phase of point $a$ and $b$ in Fig.~\ref{fig:Phase_alignment_structure}, respectively.

	Therefore, the problem \eqref{e_projection} is equivalent to the problem
	\begin{equation} \label{e_varphi_proj}
		\min_{\varphi}~\sqrt{|\tilde e|^2 + 1 - 2|\tilde e|{\rm cos}(\varphi) }~~{\rm s.t.}~\varphi \in [0,2\pi).
	\end{equation}
	Then, we have $\frac{\partial l_{ab}(\varphi)}{\partial \varphi} = \frac{ 2 |\widetilde e| {\rm sin}(\varphi) }{|\widetilde e|^2 + 1 - 2|\widetilde e| {\rm cos}(\varphi) } \geq 0$ for $\varphi\in [0,2\pi)$.
	Therefore, the optimal solution of the problem \eqref{e_varphi_proj} can be obtained when $\varphi = 0$.
	
	The proof is completed. $\hfill\blacksquare$
	\vspace{-0.3cm}

\end{appendices}

%%%%%%%%%% References %%%%%%%%%%%%%%
%\clearpage
\bibliographystyle{IEEEtran}
\bibliography{refs}

\begin{thebibliography}{10}
\providecommand{\url}[1]{#1}
\csname url@samestyle\endcsname
\providecommand{\newblock}{\relax}
\providecommand{\bibinfo}[2]{#2}
\providecommand{\BIBentrySTDinterwordspacing}{\spaceskip=0pt\relax}
\providecommand{\BIBentryALTinterwordstretchfactor}{4}
\providecommand{\BIBentryALTinterwordspacing}{\spaceskip=\fontdimen2\font plus
\BIBentryALTinterwordstretchfactor\fontdimen3\font minus
  \fontdimen4\font\relax}
\providecommand{\BIBforeignlanguage}[2]{{%
\expandafter\ifx\csname l@#1\endcsname\relax
\typeout{** WARNING: IEEEtran.bst: No hyphenation pattern has been}%
\typeout{** loaded for the language `#1'. Using the pattern for}%
\typeout{** the default language instead.}%
\else
\language=\csname l@#1\endcsname
\fi
#2}}
\providecommand{\BIBdecl}{\relax}
\BIBdecl

\bibitem{wu2020risreview}
Q.~Wu and R.~Zhang, ``Towards smart and reconfigurable environment: Intelligent
  reflecting surface aided wireless network,'' \emph{IEEE Commun. Mag.},
  vol.~58, no.~1, pp. 106--112, Jan. 2020.

\bibitem{Basar2024RIS}
E.~Basar, G.~C. Alexandropoulos, Y.~Liu, Q.~Wu, S.~Jin, C.~Yuen, O.~A. Dobre,
  and R.~Schober, ``Reconfigurable intelligent surfaces for {6G}: Emerging
  hardware architectures, applications, and open challenges,'' \emph{IEEE
  Trans. Veh. Technol.}, vol.~19, no.~3, pp. 27--47, Sept. 2024.

\bibitem{Wu2019IRS}
Q.~Wu and R.~Zhang, ``Intelligent reflecting surface enhanced wireless network
  via joint active and passive beamforming,'' \emph{IEEE Trans. Wireless
  Commun.}, vol.~18, no.~11, pp. 5394--5409, Nov. 2019.

\bibitem{Ahmed2025ActiveRIS}
M.~Ahmed, S.~Raza, A.~A. Soofi, F.~Khan, W.~U. Khan, S.~Z.~U. Abideen, F.~Xu,
  and Z.~Han, ``Active reconfigurable intelligent surfaces: Expanding the
  frontiers of wireless communication-a survey,'' \emph{IEEE Commun. Surv.
  Tutor.}, vol.~27, no.~2, pp. 839--869, 2025.

\bibitem{zhang2023secrecy}
Q.~Zhang, J.~Liu, Z.~Gao, Z.~Li, Z.~Peng, Z.~Dong, and H.~Xu, ``Robust
  beamforming design for {RIS}-aided {NOMA} secure networks with transceiver
  hardware impairments,'' \emph{IEEE Trans. Commun.}, vol.~71, no.~6, pp.
  3637--3649, Jun. 2023.

\bibitem{Yuan2023NonTerrestrial}
J.~Yuan, G.~Chen, M.~Wen, R.~Tafazolli, and E.~Panayirci, ``Secure transmission
  for {THz}-empowered {RIS}-assisted non-terrestrial networks,'' \emph{IEEE
  Trans. Veh. Technol.}, vol.~72, no.~5, pp. 5989--6000, May 2023.

\bibitem{Zheng2023ARIS}
S.~Zheng, B.~Lv, T.~Zhang, Y.~Xu, G.~Chen, R.~Wang, and P.~C. Ching, ``On {DoF}
  of active {RIS}-assisted {MIMO} interference channel with arbitrary antenna
  configurations: When will {RIS} help?'' \emph{IEEE Trans. Veh. Technol.},
  vol.~72, no.~12, pp. 16\,828--16\,833, Dec. 2023.

\bibitem{Yu2024RIS_ISAC}
Z.~Yu, H.~Ren, C.~Pan, G.~Zhou, B.~Wang, M.~Dong, and J.~Wang, ``Active
  {RIS}-aided {ISAC} systems: Beamforming design and performance analysis,''
  \emph{IEEE Trans. Commun.}, vol.~72, no.~3, pp. 1578--1595, Mar. 2024.

\bibitem{Zhang2023ActiveRIS}
Z.~Zhang, L.~Dai, X.~Chen, C.~Liu, F.~Yang, R.~Schober, and H.~V. Poor,
  ``Active {RIS} vs. passive {RIS}: Which will prevail in {6G}?'' \emph{IEEE
  Trans. Commun.}, vol.~71, no.~3, p. 1707–1725, Mar. 2023.

\bibitem{Zhang2025BD_RIS}
Q.~Zhang, G.~Luo, Z.~Dong, F.~Sun, X.~Wang, and J.~Liu, ``Beyond-diagonal
  reconfigurable intelligent surface enhanced {NOMA} systems,'' \emph{IEEE
  Wireless Commun. Lett.}, vol.~14, no.~1, pp. 118--122, Jan. 2025.

\bibitem{bjornson2021rischannel}
E.~Björnson and L.~Sanguinetti, ``Rayleigh fading modeling and channel
  hardening for reconfigurable intelligent surfaces,'' \emph{IEEE Wireless
  Commun. Lett.}, vol.~10, no.~4, p. 830–834, Apr. 2021.

\bibitem{Yang2024XL_RIS}
S.~Yang, C.~Xie, W.~Lyu, B.~Ning, Z.~Zhang, and C.~Yuen, ``Near-field channel
  estimation for extremely large-scale reconfigurable intelligent surface
  ({XL-RIS})-aided wideband mmwave systems,'' \emph{IEEE J. Sel. Areas
  Commun.}, vol.~42, no.~6, pp. 1567--1582, Jun. 2024.

\bibitem{Yang2023CE_XL_RIS}
S.~Yang, W.~Lyu, Z.~Hu, Z.~Zhang, and C.~Yuen, ``Channel estimation for
  near-field {XL-RIS}-aided mmwave hybrid beamforming architectures,''
  \emph{IEEE Trans. Veh. Technol.}, vol.~72, no.~8, pp. 11\,029--11\,034, Aug.
  2023.

\bibitem{Gong2024HMIMO}
T.~Gong, P.~Gavriilidis, R.~Ji, C.~Huang, G.~C. Alexandropoulos, L.~Wei,
  Z.~Zhang, M.~Debbah, H.~V. Poor, and C.~Yuen, ``Holographic {MIMO}
  communications: Theoretical foundations, enabling technologies, and future
  directions,'' \emph{IEEE Commun. Surv. Tutor.}, vol.~26, no.~1, pp. 196--257,
  Firstquarter 2024.

\bibitem{Gong2024NFMIMO}
T.~Gong, L.~Wei, C.~Huang, G.~C. Alexandropoulos, M.~Debbah, and C.~Yuen,
  ``Near-field channel modeling for holographic {MIMO} communications,''
  \emph{IEEE Wireless Commun.}, vol.~31, no.~3, pp. 108--116, Jun. 2024.

\bibitem{An2024NFComm}
J.~An, C.~Yuen, L.~Dai, M.~D. Renzo, M.~Debbah, and L.~Hanzo, ``Near-field
  communications: Research advances, potential, and challenges,'' \emph{IEEE
  Wireless Commun.}, vol.~31, no.~3, pp. 100--107, Jun. 2024.

\bibitem{Cui2023CommunMag}
M.~Cui, Z.~Wu, Y.~Lu, X.~Wei, and L.~Dai, ``Near-field {MIMO} communications
  for {6G}: Fundamentals, challenges, potentials, and future directions,''
  \emph{IEEE Commun. Mag.}, vol.~61, no.~1, pp. 40--46, Jan. 2023.

\bibitem{Liu2023NearField}
Y.~Liu, Z.~Wang, J.~Xu, C.~Ouyang, X.~Mu, and R.~Schober, ``Near-field
  communications: A tutorial review,'' \emph{IEEE Open J. Commun. Soc.},
  vol.~4, pp. 1999--2049, 2023.

\bibitem{Lu2024XL_MIMO}
Y.~Lu, Z.~Zhang, and L.~Dai, ``Hierarchical beam training for extremely
  large-scale {MIMO}: From far-field to near-field,'' \emph{IEEE Trans.
  Commun.}, vol.~72, no.~4, pp. 2247--2259, Apr. 2024.

\bibitem{Wei2022Chinacomm}
X.~Wei, L.~Dai, Y.~Zhao, G.~Yu, and X.~Duan, ``Codebook design and beam
  training for extremely large-scale {RIS}: Far-field or near-field?''
  \emph{China Commun.}, vol.~19, no.~6, pp. 193--204, Jun. 2022.

\bibitem{Shen2023Multi_beam}
D.~Shen, L.~Dai, X.~Su, and S.~Suo, ``Multi-beam design for near-field
  extremely large-scale {RIS}-aided wireless communications,'' \emph{IEEE
  Trans. Green Commun. Netw.}, vol.~7, no.~3, pp. 1542--1553, Sept. 2023.

\bibitem{Zhang2022FastBeamTraining}
Y.~Zhang, X.~Wu, and C.~You, ``Fast near-field beam training for extremely
  large-scale array,'' \emph{IEEE Wireless Commun. Lett.}, vol.~11, no.~12, pp.
  2625--2629, Dec. 2022.

\bibitem{Lv2024codebook}
S.~Lv, Y.~Liu, X.~Xu, A.~Nallanathan, and A.~L. Swindlehurst, ``{RIS}-aided
  near-field {MIMO} communications: Codebook and beam training design,''
  \emph{IEEE Trans. Wireless Commun.}, vol.~23, no.~9, pp. 12\,531--12\,546,
  Sept. 2024.

\bibitem{Ke2020CS}
M.~Ke, Z.~Gao, Y.~Wu, X.~Gao, and R.~Schober, ``Compressive sensing-based
  adaptive active user detection and channel estimation: Massive access meets
  massive {MIMO},'' \emph{IEEE Trans. Signal Process.}, vol.~68, pp. 764--779,
  2020.

\bibitem{Lim2020EfficientBeamT}
S.~H. Lim, S.~Kim, B.~Shim, and J.~W. Choi, ``Efficient beam training and
  sparse channel estimation for millimeter wave communications under
  mobility,'' \emph{IEEE Trans. Commun.}, vol.~68, no.~10, pp. 6583--6596, Oct.
  2020.

\bibitem{Wei2022XL_MIMO}
X.~Wei and L.~Dai, ``Channel estimation for extremely large-scale massive
  {MIMO}: Far-field, near-field, or hybrid-field?'' \emph{IEEE Commun. Lett.},
  vol.~26, no.~1, pp. 177--181, Jan. 2022.

\bibitem{Shi2011WMMSE}
Q.~Shi, M.~Razaviyayn, Z.~Q. Luo, and C.~He, ``An iteratively weighted {MMSE}
  approach to distributed sum-utility maximization for a {MIMO} interfering
  broadcast channel,'' \emph{IEEE Trans. Signal Process.}, vol.~59, no.~9, pp.
  4331--4340, Sept. 2011.

\bibitem{Xiao2016HCodebook_Criteria}
Z.~Xiao, T.~He, P.~Xia, and {X. -G. Xia}, ``Hierarchical codebook design for
  beamforming training in millimeter-wave communication,'' \emph{IEEE Trans.
  Wireless Commun.}, vol.~15, no.~5, pp. 3380--3392, May 2016.

\bibitem{Zhang2024PracticalRIS}
Q.~Zhang, J.~Liu, H.~Tang, Z.~Dong, and Y.~Li, ``Practical {RIS}-aided
  multiuser communications with imperfect {CSI}: Practical model, amplitude
  feedback, and beamforming optimization,'' \emph{IEEE Trans. Wireless
  Commun.}, vol.~23, no.~10, pp. 15\,245--15\,260, Oct. 2024.

\bibitem{Guo2020ProxLinear}
H.~Guo, {Y. -C. Liang}, J.~Chen, and E.~G. Larsson, ``Weighted sum-rate
  maximization for reconfigurable intelligent surface aided wireless
  networks,'' \emph{IEEE Trans. Wireless Commun.}, vol.~19, no.~5, pp.
  3064--3076, May 2020.

\bibitem{Nadeem2022MaxMinSINR}
{Q. -U. -A. Nadeem, A. Zappone, and A. Chaaban}, ``Achievable rate analysis and
  max-min sinr optimization in intelligent reflecting surface assisted
  cell-free mimo uplink,'' \emph{IEEE Open J. Commun. Soc.}, vol.~3, pp.
  1295--1322, 2022.

\bibitem{Shen2018FP}
K.~Shen and W.~Yu, ``Fractional programming for communication systems—part
  {I}: Power control and beamforming,'' \emph{IEEE Trans. Signal Process.},
  vol.~66, no.~10, pp. 2616--2630, May 2018.

\bibitem{Luo2023SDR}
Z.~Luo, W.~Ma, A.~M. So, Y.~Ye, and S.~Zhang, ``Semidefinite relaxation of
  quadratic optimization problems,'' \emph{IEEE Signal Process. Mag.}, vol.~27,
  no.~3, p. 20–34, May 2010.

\bibitem{Han2020NF_Channel}
Y.~Han, S.~Jin, C.~K. Wen, and X.~Ma, ``Channel estimation for extremely
  large-scale massive {MIMO} systems,'' \emph{IEEE Wireless Commun. Lett.},
  vol.~9, no.~5, pp. 633--637, 2020.

\bibitem{Zhao2024RIS_Demo}
Y.~Zhao, Y.~Feng, A.~M. Ismail, Z.~Wang, Y.~L. Guan, Y.~Guo, and C.~Yuen,
  ``2-bit ris prototyping enhancing rapid-response space-time wavefront
  manipulation for wireless communication: Experimental studies,'' \emph{IEEE
  Open J. Commun. Soc.}, vol.~5, pp. 4885--4901, 2024.

\bibitem{Shi2020PDD}
Q.~Shi and M.~Hong, ``Penalty dual decomposition method for nonsmooth nonconvex
  optimization—part {I}: Algorithms and convergence analysis,'' \emph{IEEE
  Trans. Signal Process.}, vol.~68, pp. 4108--4122, 2020.

\bibitem{Boyd2004convex}
S.~Boyd, S.~P. Boyd, and L.~Vandenberghe, \emph{Convex Optimization}.\hskip 1em
  plus 0.5em minus 0.4em\relax Cambridge, U.K.: Cambridge Univ. Press, 2004.

\bibitem{Dong2022NFChannel}
Z.~Dong and Y.~Zeng, ``Near-field spatial correlation for extremely large-scale
  array communications,'' \emph{IEEE Commun. Lett.}, vol.~26, no.~7, pp.
  1534--1538, Jul. 2022.

\bibitem{Noh2017FFHL}
S.~Noh, M.~D. Zoltowski, and D.~J. Love, ``Multi-resolution codebook and
  adaptive beamforming sequence design for millimeter wave beam alignment,''
  \emph{IEEE Trans. Wireless Commun.}, vol.~16, no.~9, p. 5689–5701, Sept.
  2017.

\bibitem{Lee2016FFES}
J.~Lee, {G.-T. Gil}, and Y.~H. Lee, ``Channel estimation via orthogonal
  matching pursuit for hybrid {MIMO} systems in millimeter wave
  communications,'' \emph{IEEE Trans. Commun.}, vol.~64, no.~6, p. 2370–2386,
  Jun. 2016.

\end{thebibliography}

\end{document}